\documentclass[preprint,preprintnumbers,amsmath,amssymb,floatfix,nofootinbib,superscriptaddress]{revtex4}
\pdfoutput=1

\usepackage[dvips]{graphicx}
\usepackage{url}
\usepackage{subfigure}
\usepackage[usenames,dvipsnames]{color}
\usepackage{xspace}
\usepackage{float}
\usepackage[pagebackref]{hyperref}
\usepackage[normalem]{ulem}
\usepackage{setspace}

\def\be{\begin{equation}}
\def\ee{\end{equation}}
\def\ba{\begin{eqnarray}}
\def\ea{\end{eqnarray}}

\def\ln{\mathrm{ln}}

\def\msbar{\overline{\mathrm{MS}}}

\def\MCFM{{\tt MCFM}}

\newcommand{\nlox}{\texttt{NLOX}\xspace}
\newcommand{\python}{\texttt{Python}\xspace}
\newcommand{\qgraf}{\texttt{QGRAF}\xspace}
\newcommand{\form}{\texttt{FORM}\xspace}
\newcommand{\cpp}{\texttt{C++}\xspace}
\newcommand{\tred}{\texttt{tred}\xspace}

\newcommand{\order}[1]{O(#1)}
\newcommand{\eps}{\epsilon}

\newcommand{\ir}{\mathrm{ir}}
\newcommand{\uv}{\mathrm{uv}}

\begin{document}


\title{Electroweak and QCD corrections to $Z$-boson production with 
one $b$ jet in a massive 5 Flavor Scheme}

\author{D.~Figueroa}
\email{daf14f@my.fsu.edu}
\affiliation{Physics Department, Florida State University,
Tallahassee, FL 32306-4350, U.S.A.}
\author{S.~Honeywell}
\email{sjh07@hep.fsu.edu}
\affiliation{Physics Department, Florida State University,
Tallahassee, FL 32306-4350, U.S.A.}
\author{S.~Quackenbush}
\email{squackenbush@hep.fsu.edu}
\affiliation{Physics Department, Florida State University,
Tallahassee, FL 32306-4350, U.S.A.}
\author{L.~Reina}
\email{reina@hep.fsu.edu}
\affiliation{Physics Department, Florida State University,
Tallahassee, FL 32306-4350, U.S.A.}
\author{C.~Reuschle}
\email{christian.reuschle@thep.lu.se, creuschle@hep.fsu.edu}
\affiliation{Physics Department, Florida State University,
Tallahassee, FL 32306-4350, U.S.A.}
\affiliation{Department of Astronomy and Theoretical Physics, 
Lund University, SE-223 62 Lund, Sweden}
\author{D.~Wackeroth}
\email{dw24@buffalo.edu, dow@ubpheno.physics.buffalo.edu}
\affiliation{Department of Physics, SUNY at Buffalo, 
Buffalo, NY 14260-1500, U.S.A.\\\vspace{5ex}}


\begin{abstract}
We compute the $O(\alpha_s \alpha^2)$ and $O(\alpha_s^2 \alpha)$ 
contributions to the production cross section of a $Z$ boson with one 
$b$ jet at the Large Hadron Collider (LHC), and study their
phenomenological relevance for LHC physics. The accurate prediction of 
hadronic $Z+b$-jet production is needed to control a background that 
greatly affects both the measurement of Higgs-boson properties and 
searches of new physics at the LHC. At the same time it could enable 
the first precise measurement of the $b$-quark parton distribution 
function. In this context $b$-quark mass effects become relevant and 
need to be studied with care, both at the level of the hard process 
and at the level of the initial- and final-state parton evolution. It 
is the aim of this paper to explore some of these issues in the 
framework of a massive 5 Flavor Scheme and to assess the need for both 
the inclusion of electroweak corrections, in addition to QCD 
corrections, and $b$-quark mass effects in the prediction of total and 
differential cross sections for hadronic $Z+b$-jet production.
\end{abstract}



\maketitle

\section{Introduction}
\label{sec:introduction}

The production of a $Z$ boson with one or more $b$ jets plays a very 
important role in the physics program of the Large Hadron Collider 
(LHC) both for direct searches of physics beyond the Standard Model 
(SM) and for precision measurements of SM processes that could reveal 
deviations induced by new physics beyond the direct
reach of the LHC. Dedicated experimental studies from the LHC, which 
upgraded previous Tevatron results~\cite{Aaltonen:2008mt,
Abazov:2013uza}, have been published during Run~I~\cite{Aad:2011jn,
Aad:2014dvb,Chatrchyan:2012vr,Chatrchyan:2013zja,Chatrchyan:2014dha,
Khachatryan:2016iob,
Aaij:2014gta}, and we look forward to results obtained at 
higher center-of-mass (c.m.) energy and with much higher statistics 
during Run II.

$Z+b$ jet(s) is an important reducible and irreducible background to
several SM and beyond the SM (BSM) processes involving $Z$ bosons and
jets.  Indeed, signals of physics beyond the SM will likely emerge
from signatures containing heavy SM particles, such as $Z$ and $W$
bosons, together with $t$ and $b$ quarks. On the other hand, in the
case, for instance, of SM Higgs production, $Z+b$ jet(s) is the
dominant background to the $ZH$ associated production mode, with the
Higgs decaying into a $b\bar{b}$ pair. The theoretical accuracy of
the prediction of $Z+b$ jets will therefore affect the precision on
measurements of Higgs-boson couplings reached at the LHC.

Besides searches for new physics, the interest in a precision 
measurement of $Z+b$-jet production is also motivated by the 
possibility of obtaining the first direct measurement of the $b$-quark
parton distribution function (PDF). Indeed, if one assumes a non-zero
$b$-quark PDF, i.e. if one works in a 5 Flavor Scheme (5FS), the
production of a $Z$ boson with one $b$ jet proceeds dominantly via
$bg\rightarrow Zb$. If this is very natural at energy scales much
larger than $m_b$, it becomes less justified at lower energies where
working in a 4 Flavor Scheme (4FS) may also be appropriate, and the
channels $q\bar{q}\rightarrow Zb\bar{b}$ and $gg\rightarrow
Zb\bar{b}$ are the main channels for $Z$ production with both one and 
two $b$ jets.  In the last few years a lot of theoretical activity has 
gone into clarifying the interplay between the two approaches~\cite{
Campbell:2008hh,Caola:2011pz,Maltoni:2012pa,Forte:2015hba,Lim:2016wjo,
Forte:2016sja,Bonvini:2015pxa,Bonvini:2016fgf,Krauss:2016orf} (for a 
review see also Ref.~\cite{Cordero:2015sba}), and times are now mature 
to develop a dedicated experimental program aimed at the measurement 
of the $b$-quark parton density via processes like $Z+b$ jets, or 
$\gamma+b$ jets, which provides analogous yet complementary
information.

As already proven by the incredibly successful physics program of
Run~I of the LHC, both the intricacy of new physics searches and the
challenge of SM precision measurements require the improvement of
the accuracy of theoretical predictions to the percent level. If the 
knowledge of the first order(s) of QCD corrections was mandatory for 
Run~I, electroweak (EW) corrections will also become important at the
energies of Run~II. At the same time, a more accurate assessment of
the theoretical uncertainties intrinsic to sophisticated Monte Carlo
tools used to match higher-order QCD/EW calculations to parton-shower
generators is clearly needed. Among others, the question of properly
including heavy-quark mass effects ($b$-quark mass effects in our 
case), both at the level of the hard-scattering matrix element and at 
the level of the PDF, should be carefully considered, in particular 
for $b$-quark initiated processes. Given its relevance for the physics 
of Run~II, $Z+b$-jet production offers a particularly interesting 
prototype case to be considered in this context.

With this in mind, we consider in this paper the production of a $Z$
boson with one $b$ jet as generated at Leading Order (LO) via the
tree-level $O(\alpha_s\alpha)$ process $bg\rightarrow Zb$.  The first
order of QCD corrections, i.e. the $O(\alpha_s$) or Next-to-Leading
Order (NLO) QCD corrections to this process, have been calculated for
the first time in Ref.~\cite{Campbell:2003dd} and implemented in
\MCFM~\cite{mcfm8}, assuming a massless $b$ quark.  To improve the
precision of the NLO QCD theoretical predictions one could either add
the second order (i.e. the next-to-next-to-leading order or NNLO) QCD
corrections or the first order (i.e. the NLO) EW corrections. Adding
NNLO QCD corrections should certainly stabilize the cross section by
reducing the dependence on the renormalization and factorization
scales, while adding NLO EW corrections could add a few percent to
total rates and have a visible impact on high-energy tails of
distributions. NNLO QCD corrections to $Z$+jets have been presented in
Refs.~\cite{Ridder:2015dxa,Boughezal:2015ded} and, with due care, they
could be used in the future to extract the NNLO QCD prediction for
$Z+b$-jet for the case of a massless $b$ quark.  The $O(\alpha_s)$
corrections to $q\bar{q},gg\rightarrow Zb\bar{b}$ for a massive $b$
quark have also been
calculated~\cite{FebresCordero:2008ci,Cordero:2009kv}. They represent
an important component of the NNLO QCD corrections for massive $b$
quarks and indicate a sensible reduction of the overall scale
dependence for $Z+b$ jet production.  Obtaining NNLO QCD corrections
for a massive $b$ quark is a more challenging endeavor and should be
considered only after the inclusion of massive initial-state partons,
some issues of which we discuss in this paper. NLO EW corrections to
hadronic $Z+j$ production have been presented in
Refs.~\cite{Kuhn:2005az,Denner:2010mu,Denner:2011vu,Hollik:2015pja},
and combined NLO QCD+EW corrections in Ref.~\cite{Kallweit:2015dum}.
The first order of EW corrections to $Z+b$-jet production via
$bg\rightarrow Zb$, i.e. the $O(\alpha_s\alpha^2)$ term in the
perturbative expansion of the cross section for $Z+b$-jet production,
is the main subject of this paper, where we also study the effect of
considering an initial-state massive $b$ quark and discuss the
interplay with the definition of the corresponding $b$-quark PDF. This
can be considered as the first building block of a more general
program that will have to connect both 4FS and 5FS calculations,
including both QCD and EW corrections.  Since the EW corrections to
$bg\rightarrow Zb$ represent a well defined set of corrections, in a
well defined flavor scheme, this calculation allows to estimate in a
consistent way the impact of EW corrections on $Z+b$-jet production
through the process that also most affects the measurement of the
$b$-quark PDF.

Furthermore, working with the idea of implementing a 5FS calculation
in a parton-shower Monte Carlo event generator, we treat the $b$ quark
as massive also in the initial state, since this is necessary in order
to properly implement the backward evolution of final-state massive
$b$ quarks. Indeed, if it is customary to treat initial-state $b$
quarks as massless (traditional 5FS), this is nevertheless just a
simplification of the calculation, not a
requirement~\cite{Collins:1998rz,Aivazis:1993pi,Kramer:2000hn}.  More
to the point, this approximation does not lend itself well to the
implementation of methods, like phase-space slicing methods or
subtraction methods, which evaluate higher-order real radiative
corrections with the help of auxiliary terms that algorithmically
approximate the real-emission contributions in the soft/collinear
regions starting from the corresponding Born processes.  For example,
a real emission subtraction term to the real-emission process
$gg\rightarrow Zb\bar{b}$, with massive $b$-quarks, cannot be
generated in a kinematically consistent way from the corresponding
Born process $bg\rightarrow Zb$, with massless $b$-quarks, by
convoluting with the splitting function for $g\rightarrow b\bar{b}$.
Since it is, for example, at the core of the implementation of 5FS
processes in Monte Carlo event generators that match NLO cross
sections to parton showers, this issue has recently attracted some
attention and studies aimed at introducing what has been dubbed as
\textit{massive 5FS} (m5FS\,%
\footnote{So far we have used 5FS to denote a generic scheme with 5 
  active flavors. In the following we will use 5FS to denote the case 
  in which the $b$ quark is considered massless, and m5FS to denote 
  the case in which the $b$ quark is considered massive.}%
) have appeared~\cite{Krauss:2017wmx}.  Hence, in view of future 
developments in Monte Carlo event generators, we consistently develop 
the calculation of both the first order of QCD and EW corrections with 
a massive $b$ quark. This incidentally also implies that we extend the 
existing NLO QCD calculation~\cite{Campbell:2003dd} to the case of a 
massive $b$ quark.

In this paper, both QCD and EW virtual corrections have been obtained
through the \nlox one-loop provider~\cite{NLOX}, as well as by
independent in-house codes. The corresponding real corrections have
been computed, in both the massless and massive cases, via independent
codes using phase-space slicing, as well as an implementation of the
dipole subtraction method based on the formalism of
Ref.~\cite{Dittmaier:1999mb}, extended to QED radiation off massive
quarks (in both initial and final state).  More details will be
presented in Section~\ref{sec:calculation}.  Having obtained both the
$O(\alpha_s)$ and the $O(\alpha)$ corrections to $bg\rightarrow Zb$
including full $b$-quark mass effects, we can assess: 1) the impact of
mass effects on the fixed-order total cross section and distributions
by comparing NLO QCD cross sections with massless and massive $b$
quarks, and 2) the relative impact of QCD and EW corrections on
fixed-order total cross sections and distributions by comparing the
$O(\alpha_s^2\alpha)$ and $O(\alpha_s\alpha^2)$ cross sections with
massive $b$ quarks.  Indeed, independently of the necessity of
introducing massive initial-state partons for automated
implementations of NLO QCD and EW corrections, it is also clear that
mass effects and EW corrections can be of the same order, although
they typically affect physical observables in different kinematical
regions. Even if small, they both need to be accounted for when one
aims for percent-level precision predictions over a broad kinematical
range.

In the remainder of this paper we systematically review in
Section~\ref{sec:calculation} the relevant technical details of the
calculation and present results in Section~\ref{sec:results} where we
also assess the impact of such corrections and discuss the need for
future improvements.  Conclusions and suggestions for future
developments are presented in Section~\ref{sec:conclusions}.

\section{Details of the calculation}
\label{sec:calculation}

\begin{figure}
\includegraphics[scale=1]{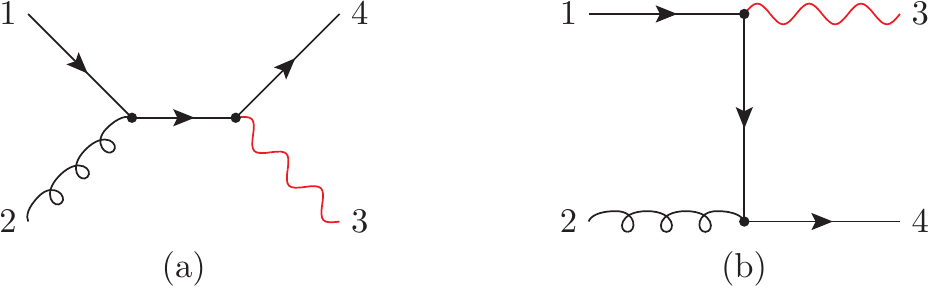}
\caption{\label{fig:tree_g1e1}\small Tree-level ($s$- and $t$-channel) 
Feynman diagrams for $b(1)g(2)\rightarrow Z(3)b(4)$.}
\end{figure}

We write the hadronic cross section, $\sigma$, for $Z+b$-jet
production at the LHC as follows:
\begin{equation}
\label{eq:sigma_ij}
\sigma=\sum\limits_{i,j}\alpha_s^i\alpha^j\sigma^{(i,j)}\,,
\end{equation}
where $i$ and $j$, with $i+j\geq 2$ and $j\geq 1$, refer to the 
coupling order of the partonic cross section, and $\sigma^{(i,j)}$ 
denotes the term in the perturbative expansion of the cross section 
that is proportional to $\alpha_s^i\alpha^j$ (where 
$\alpha_s=g_s^2/(4\pi)$ and $\alpha=e^2/(4\pi)$, $g_s$ and $e$ being 
the QCD and QED coupling constants, respectively). In a scheme with 5 
active flavors (such as the 5FS or m5FS),  the set of "lowest-order" 
contributions with $i+j=2$  consists of all sets of tree-level 
diagrams that satisfy such relation, i.e. $\sigma^{(1,1)}$ corresponds 
to the tree-level contributions to 
$bg\rightarrow Zb$ and 
$\sigma^{(0,2)}$ to the tree-level contributions to
$b\gamma\rightarrow Zb$\,%
\footnote{Note that the set of all possible "lowest-order" 
  contributions is made of contributions of different coupling-power 
  combinations and that those also arise from different sets of 
  initial-state particles. The same holds for the set of all possible 
  "next-to-lowest-order" contributions.}%
. It is clear that the calculation of the cross section also includes 
the corresponding $\bar{b}g$-initiated and $\bar{b}\gamma$-initiated
processes, even if it is not explicitly repeated throughout the paper. 
Both processes consist of $s$- and 
$t$-channel contributions, as illustrated in Fig.~\ref{fig:tree_g1e1} 
for the case of $bg\rightarrow Zb$ (the corresponding diagrams for
$b\gamma\rightarrow Zb$ are obtained by replacing a gluon with a
photon in both diagrams). 

\begin{figure}
\hspace{-4.5ex}\includegraphics[scale=1]{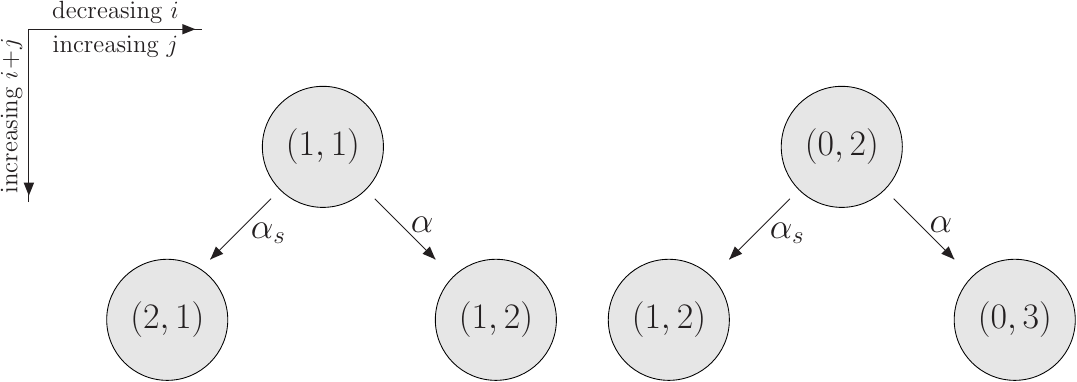}
\caption{\label{fig:couplingpowerflow} Coupling-power flow chart for 
$Z+b$-jet production at the level of the partonic cross section. The 
notation $(i,j)$ corresponds to the coupling-power combination 
$\alpha_s^i\alpha^j$. From left to right we have increasing $i$ / 
decreasing $j$ in steps of 1. From top to bottom the total order $i+j$ 
increases in steps of 1. The upper row depicts all possible 
coupling-power combinations for the "lowest-order" contributions, with 
$i+j=2$, while the lower row depicts all possible coupling-power 
combinations for the higher-order corrections of one order higher, 
with $i+j=3$. Note that the $(1,2)$ contribution is depicted twice, as 
it originates from two different "lowest-order" contributions with 
different sets of initial-state particles.}
\end{figure}

The set of "next-to-lowest-order" contributions consists of the 
$O(\alpha_s)$ and $O(\alpha)$ corrections to the set of "lowest-order"
contributions, as depicted in Fig.~\ref{fig:couplingpowerflow}.
The $O(\alpha_s^2\alpha)$ term, $\sigma^{(2,1)}$, corresponds to the
$O(\alpha_s)$ corrections to $\sigma^{(1,1)}$.  It was first computed
in Ref.~\cite{Campbell:2003dd} for massless $b$ quarks, and the
corresponding calculation is implemented in \MCFM~\cite{mcfm8}. We
have reproduced it independently in this paper, where we also extend
it to the case of a massive $b$ quark in order to consistently compare
the impact of NLO QCD and EW corrections, and in order to assess
non-zero $b$-quark mass effects by comparing massive and massless NLO
QCD results.  The $O(\alpha_s\alpha^2)$ term, $\sigma^{(1,2)}$, is
also presented for the first time in this paper and is indeed the main
focus of our study.  Such term originates from both the $O(\alpha)$
corrections to $\sigma^{(1,1)}$ ($bg\rightarrow Zb$) and the
$O(\alpha_s)$ corrections to $\sigma^{(0,2)}$
($b\gamma\rightarrow Zb$), as depicted in
Fig.~\ref{fig:couplingpowerflow}.  It is entirely dominated by the
$O(\alpha)$ corrections to the $bg\rightarrow Zb$ tree-level process,
since the cross section for $b\gamma\rightarrow Zb$ is much smaller
due to the smallness of the photon parton density in the initial-state
protons\,%
\footnote{Using the setup described in Section~\ref{sec:results}, one
  finds that
  $\sigma_{\mbox{\tiny LO}}=\alpha_s\alpha\sigma^{(1,1)}\simeq
  376$~pb,
  while the cross section for the ``lowest-order'' photon-induced
  process is three orders of magnitude smaller,
  $\alpha^2\sigma^{(0,2)}\simeq 0.1$~pb, and one order of magnitude
  smaller than $\alpha_s\alpha^2\,\sigma^{(1,2)}\simeq 5$~pb.}%
.  Hence in our study we will only consider the $O(\alpha)$
corrections to $bg\rightarrow Zb$.  Also the $O(\alpha^3)$ term,
$\sigma^{(0,3)}$, is entirely negligible and will not be considered
here.  We will thus from here on simply speak of $\sigma^{(1,1)}$ as
the LO contribution, and define
\begin{equation}
\label{eq:sigma_lo}
\alpha_s\alpha\,\sigma^{(1,1)}+
\alpha^2\sigma^{(0,2)}
\approx
\alpha_s\alpha\,\sigma^{(1,1)}
\equiv\sigma_{\mbox{\tiny LO}}\,,
\end{equation}
and of 
$\sigma^{(2,1)}$ and $\sigma^{(1,2)}$ as the $O(\alpha_s)$ and 
$O(\alpha)$ corrections, or simply NLO QCD and EW corrections, 
respectively, and define the corresponding NLO QCD and NLO EW cross 
sections as 
\begin{eqnarray}
\label{eq:sigma_nlo}
\sigma^{\mbox{\tiny QCD}}_{\mbox{\tiny NLO}} & \equiv&\sigma_{\mbox{\tiny LO}}
+ \alpha_s^2\alpha\,\sigma^{(2,1)}, \nonumber \\
\sigma^{\mbox{\tiny EW}}_{\mbox{\tiny NLO}} & \equiv& \sigma_{\mbox{\tiny LO}} 
+ \alpha_s\alpha^2\,\sigma^{(1,2)} \; .
\end{eqnarray}

In order to implement both the NLO EW and NLO QCD cross sections
in the m5FS, the full $b$-quark mass
dependence has been retained both in the hard-scattering matrix elements
and in the phase-space integration. We therefore have defined the
initial-state parton-level kinematics of $b(p_b)+g(p_g)\rightarrow Zb$
as follows~\cite{Nagy:2014oqa}:
\begin{eqnarray}
\label{eq:bg-massive-kinematics}
p_b^\mu&=&\frac{\sqrt{s}}{2}\left(
x_1+\frac{m_b^2}{x_1 s},0,0,x_1-\frac{m_b^2}{x_1 s}\right)
\,\,\,\,\,\mbox{with}\,\,\,\,\, 
\frac{m_b}{\sqrt{s}}\le x_1\le\frac{1}{2}\left(1+\sqrt{1-\frac{4m_b^2}{s}}
\right)\,,\\
p_g^\mu&=&\frac{\sqrt{s}}{2}\left(
x_2,0,0,-x_2\right)
\,\,\,\,\,\mbox{with}\,\,\,\,\, 0\le x_2\le 1\,,
\nonumber
\end{eqnarray}
where $\sqrt{s}$ is the hadronic c.m. energy, $x_1$ and $x_2$ are the
longitudinal fractions of the corresponding hadron momenta carried by
each parton. Their ranges assure compatibility with the parton-level
kinematics in the hadronic c.m.  frame, where we impose that
$m_b\le p_b^0\le\frac{\sqrt{s}}{2}$ and
$0\le p_g^0\le \frac{\sqrt{s}}{2}$. Of course, the symmetric case in
which $x_1$ and $x_2$ are exchanged is also considered, and, due to
the massiveness of the initial-state $b$ quark, care must be taken to
keep the two kinematic cases separate when convoluting the partonic
cross section with the PDF to obtain the hadronic cross section. In
the case of $b\bar{b}\rightarrow Zb\bar{b}$ with an initial-state
massive $b$, the initial-state kinematic configuration
becomes~\cite{Nagy:2014oqa}
\begin{align}
\label{eq:bb-Zbb-massive-kinematics}
&p_b=\frac{\sqrt{s}}{2}\left(
x_1+\frac{m_b^2}{x_1 s},0,0,x_1-\frac{m_b^2}{x_1 s}\right) \,\,\,\,\text{ and }
\,\,\,\,p_{\bar{b}}=\frac{\sqrt{s}}{2}\left( x_2+\frac{m_b^2}{x_2
  s},0,0,-x_2+\frac{m_b^2}{x_2 s}\right)\\ 
&\!\!\text{ with }\,\,\,\,
\frac{m_b}{\sqrt{s}}\le
x_{1,2}\le\frac{1}{2}\left(1+\sqrt{1-\frac{4m_b^2}{s}}\right)\,.\notag
\end{align}
In the 5FS, which assumes 
a massless $b$ quark, the definition of the parton-level kinematics 
for $bg\rightarrow Zb$ is the same as for massless partons (obtained 
by setting $m_b=0$ in Eqs.~(\ref{eq:bg-massive-kinematics}) and
(\ref{eq:bb-Zbb-massive-kinematics})).

In the following we will review the essential features of the
$O(\alpha_s)$ and $O(\alpha)$ virtual and real corrections in
Sections~\ref{sec:virtual} and \ref{sec:real} respectively, and
discuss the choice of PDF subtraction terms in Section~\ref{sec:pdfsub}.

\subsection{Virtual corrections with \nlox}
\label{sec:virtual}

The virtual $O(\alpha_s)$ and $O(\alpha)$ corrections to
$\sigma_{\mbox{\tiny LO}}$, contributing to
$\sigma^{\mbox{\tiny QCD}}_{\mbox{\tiny NLO}}$ and
$\sigma^{\mbox{\tiny EW}}_{\mbox{\tiny NLO}}$ respectively, have been
produced with the one-loop provider \nlox~\cite{NLOX}, interfaced to a
selection of in-house Monte Carlo integration routines.\footnote{The
  $\order{\alpha}$ and $\order{\alpha_s}$ corrections at the
  amplitude-squared level have been provided by \nlox and
  cross-checked on the phase-space point level against several
  independent in-house codes and the one-loop provider
  \texttt{Recola}~\cite{Actis:2016mpe}. At the cross-section level
  $\texttt{NLOX}$ and $\texttt{Recola}$ have been used with a
  dedicated interface to the $\texttt{Cuba}$
  library~\cite{Hahn:2004fe}, in order to confirm the virtual
  cross-section numbers from interfacing \texttt{NLOX} and the
  in-house codes to in-house Monte Carlo integration routines.}  UV
divergences are renormalized. For what pertains to IR divergences, the
virtual IR poles are collected and their cancelation against the IR
poles arising from the $O(\alpha_s)$ and $O(\alpha)$ real corrections,
as well as from the corresponding PDF subtraction terms, is checked.

In Section~\ref{sec:virtual_nlox} we will give a brief
description of \nlox.  More details will soon be available in
Ref.~\cite{NLOX}, which will go along with a public release of the
code as well as a more general interface to publicly available Monte
Carlo integrators and event generators.
 
In Section~\ref{sec:virtual_renorm} thereafter we will
provide a description of our QCD renormalization, in order to make 
clear certain aspects that are also relevant for the discussion of
our PDF subtraction terms in Section~\ref{sec:pdfsub}.

\subsubsection{\nlox}
\label{sec:virtual_nlox}

\nlox is a new program for the automated computation of one-loop QCD
and EW corrections in the Standard Model.  A non-public predecessor of
\nlox has been available in the past, to calculate one-loop QCD
corrections to selected processes
\cite{Reina:2011mb,Reina:2012lfa}. \nlox has seen quite some progress
in recent years, and most recently partook in a technical comparison
on NLO EW automation \cite{Bendavid:2018nar}.  The current version of
the program provides fully renormalized QCD and EW one-loop
corrections in the Standard Model, for all the possible QCD+EW mixed
coupling-power combinations to a certain parton-level process up to
one-loop accuracy, including the full mass dependence on initial- and
final-state particle masses.

\nlox is based on a traditional approach of Feynman diagrams,
utilizing \qgraf \cite{Nogueira:1991ex}, \form \cite{Kuipers:2012rf},
and \python to algebraically generate \cpp code for the virtual QCD
and EW one-loop contributions to a certain process in terms of
one-loop tensor-integral coefficients.  The tensor-integral
coefficients are calculated recursively at runtime through standard
reduction methods by the \cpp library \tred, an integral part of
\nlox.  Several reduction techniques are available to \tred, many of
which are found in Refs.~\cite{Passarino:1978jh,Denner:2005nn}.  The 
scalar one-loop integrals are evaluated by either using
\texttt{OneLOop}~\cite{vanHameren:2010cp}, or
\texttt{QCDLoop}~\cite{Ellis:2007qk,Carrazza:2016gav}.

In \nlox UV and IR singularities are regularized in 
$d$-dimensional regularization (with $d=4-2\epsilon$, 
$|\epsilon|\ll1$). UV singularities are renormalized, while IR 
singularities are reported in terms of the Laurent coefficients of the 
corresponding $1/\epsilon^2$ and $1/\epsilon$ poles. The UV 
renormalization in \nlox is carried out by means of counterterm 
diagrams, which provides a flexible way to systematically include mass 
renormalization for massive propagators as well as Yukawa-type 
vertices.

The renormalization constants in terms of which the EW UV counterterms
are formulated are derived in the on-shell renormalization scheme, as
described in Ref.~\cite{Denner:1991kt}, or in the complex-mass
scheme,\footnote{Note that for $Z+b$-jet production to work in the
  on-shell approximation for a real final-state $Z$ boson we neglect
  all widths, making all masses and derived parameters real
  quantities.} as described in Ref.~\cite{Denner:2005fg}. As EW input
scheme choices \nlox provides both the $\alpha(0)$ and the $G_\mu$ EW
input schemes~\cite{Hollik:1988ii,Denner:1991kt,Dittmaier:2001ay,
  Andersen:2014efa}.  Per default the $\alpha(0)$ EW input scheme is
used.  The renormalization constants in terms of which the QCD UV
counterterms are formulated are derived in a mixed renormalization
scheme~\cite{Collins:1978wz,Nason:1987xz}: the on-shell
renormalization scheme is used for the wave-function and mass
renormalization of massive quarks, while the $\overline{\text{MS}}$
renormalization scheme is used for massless quarks and gluons, where,
however, in the latter case heavy-quark-loop contributions are
decoupled by subtracting them at zero momentum.
Since in this paper we present QCD results for both 5FS and m5FS , we
briefly discuss in Section ~\ref{sec:virtual_renorm} the
definition of those QCD counterterms that differ in the two schemes.

\subsubsection{Gluon wave-function and $\alpha_s$ QCD renormalization: 5FS vs m5FS}
\label{sec:virtual_renorm}

As discussed in Section ~\ref{sec:introduction}, the interest in
considering an initial-state massive $b$ quark and developing a m5FS
formalism does not arise from the need of treating the $b$ quark as a
heavy quark in the same way as the $t$ quark is (in which case a 4FS
would be more appropriate), but from the intent of properly matching
$b$-quark mass effects at the kinematic, hard matrix-element, and PDF level to
reduce the theoretical uncertainty in processes involving $b$
quarks/jets. The presence of a massive $b$ quark in our calculation
mainly affects the IR structure of the $b$-loop contributions, but not
the UV subtraction procedure where, in both the 5FS and the m5FS, the
only quark treated as \textit{heavy} (hence \textit{decoupled in the
  low-energy regime}) is the $t$ quark. Indeed, in our calculation we
renormalize the gluon two-point function by subtracting the contribution
of all \textit{light} quarks (including the $b$ quark) in $\msbar$ and the
contribution of the $t$ quark at zero momentum, such that only the $t$
quark is decoupled in the low-energy limit. To be more explicit, our
prescription corresponds to writing the renormalized transverse gluon
self energy as
\begin{align}
\label{eq:sigma-gluon-renormalized}
\hat{\Sigma}^{T}_{G}(p^2;\{q,Q\})
&=\Big(\;\Sigma^{T}_{G}(p^2;\{q,Q\}) - \delta Z_G(\{q,Q\})\;\Big)\\
&\hspace{-5ex}=\Big(\;\Sigma^{T}_{G}(p^2;\{q\}) \,-\,
\Sigma^{T}_{G}(p^2;\{q\})\big|_{\msbar\text{ UV pole}}\;\Big)
+\,\Big(\;\Sigma^{T}_{G}(p^2;\{Q\}) \,-\, 
\Sigma^{T}_{G}(0;\{Q\})\;\Big)\notag\,,
\end{align}
where $\{q,Q\}$ is a shorthand to denote the set $\{\{q\},\{Q\}\}$,
formed by the the sets $\{q\}$ (with dimension $n_{lf}$) and $\{Q\}$
(with dimension $n_{hf}$) of all quarks whose UV divergences from
closed quark-loop contributions are subtracted in $\msbar$ or at zero
momentum respectively. In Eq.~(\ref{eq:sigma-gluon-renormalized}),
$\Sigma^{T}_{G}(p^2;\{q\})$ contains the contributions from gluon and
ghost loops, as well as from the $\{q\}$-quark loops, while
$\Sigma^{T}_{G}(p^2;\{Q\})$ contains the contributions from the
$\{Q\}$-quark loops, where in both the 5FS and m5FS $\{Q\}=\{t\}$. This
prescription defines the gluon-field renormalization constant
$Z_G=1+\delta Z_G$, with $\delta Z_G$ given by
\begin{align}
\label{eq:dZG}
\delta Z_G(\{q,Q\})
&= -\, \Sigma^{T}_{G}(p^2;\{q\})\big|_{\msbar\text{ UV pole}}
   -\, \Sigma^{T}_{G}(0;\{Q\})\\
&\hspace{-5ex}=
-\frac{\alpha_s}{4\pi}\frac{2}{3}
S_\eps
\Big(
\,\frac{1}{\eps_\uv}\,
\Big(
(n_{lf}+n_{hf})\,2T_R
-\frac{5}{2}\,C_A 
\Big)
-\sum_{\{Q\}}2T_R\,\ln\Big(\frac{m_Q^2}{\mu^2}\Big)
\,\Big)
\notag\,,
\end{align}
and correspondingly the strong coupling renormalization constant
$Z_{g_s}=1+\delta Z_{g_s}$, with $\delta Z_{g_s}$ given by
\begin{align}
\label{eq:dZgs}
\delta Z_{g_s}(\{q,Q\})
=&
-\frac{\alpha_s}{4\pi}\frac{2}{3}\Big(\!-\!\frac{1}{2}\Big)
S_\eps
\Big(
\,\frac{1}{\eps_\uv}\,
\Big(
(n_{lf}+n_{hf})\,2T_R
-\frac{11}{2}\,C_A 
\Big)
\,-\,\sum_{\{Q\}}2T_R\,\ln\Big(\frac{m_Q^2}{\mu^2}\Big)
\,\Big)\,,
\end{align}
where in Eqs.~(\ref{eq:dZG}) and (\ref{eq:dZgs}) we have made explicit
that the poles in $\eps$ are of ultraviolet origin, $\mu$ is the 't
Hooft mass in dimensional regularization, which is typically set equal
to the renormalization scale $\mu_r$,
$S_\eps\equiv(4\pi)^\eps/\Gamma(1-\eps)$, and we have used
$C_A=N_c=3$, $C_F=(N_c^2-1)/(2N_c)$, and $T_R=1/2$.  As a result in
the m5FS the running of $\alpha_s$ is also governed by $n_f=5$
flavors, consistently with the set of PDF we choose to use (see
Sections~\ref{sec:pdfsub} and \ref{sec:results}). The only difference
introduced in considering a massive $b$ quark arises in the IR parts
of the renormalization counterterms, more specifically in the
gluon-field residue, which, given the prescription we adopted
for the gluon-field renormalization, is non-trivial and given by
\begin{align}
\label{eq:RbarG}
\bar{R}_G(\{q',Q'\},\{q,Q\})=1-\Sigma^T_G(0;\{q',Q'\})-\delta Z_G(\{q,Q\})=1+\delta\bar{R}_G\,,
\end{align}
where
\begin{align}
\label{eq:dRbarG}
\delta\bar{R}_G(\{q',Q'\},\{q,Q\})
&=
-\frac{\alpha_s}{4\pi}\frac{2}{3}
(-S_\eps)
\Big(
\;\;\,\frac{1}{\eps_\ir}\;\;
\Big(
n_{lf}'\,2T_R
-\frac{5}{2}\,C_A 
\Big)\\
&\hspace{15ex} 
+\,2T_R\,\Big(
\sum_{\{Q'\}}\ln\Big(\frac{m_{Q'}^2}{\mu^2}\Big)
-\sum_{\{Q\}}\ln\Big(\frac{m_Q^2}{\mu^2}\Big)
\Big)
\;\;\Big)
\notag\,.
\end{align}
Notice that in Eq.~(\ref{eq:dRbarG}) we have made explicit the fact
that the remaining poles in $\eps$ are of infrared origin, and we have
emphasized the difference between 5FS and m5FS by introducing a primed
notation such that: in the 5FS $\{Q^\prime\}=\{Q\}=\{t\}$ and
$n^\prime_{lf}=n_{lf}=5$, while in the m5FS $\{Q^\prime\}=\{b,t\}$,
$\{Q\}=\{t\}$, $n^\prime_{lf}=4$, and $n_{lf}=5$.

\subsection{Real corrections}
\label{sec:real}

The $O(\alpha_s)$ real corrections to $bg\rightarrow Zb$ contain both
soft and collinear singularities.
On top of gluon emission from $bg\rightarrow Zb$
($bg\rightarrow Zb+g$), they also include the
$qb (\bar{q}b)\rightarrow Zb+q(\bar{q})$, $gg\rightarrow Zb\bar{b}$,
$q\bar{q}\rightarrow Zb\bar{b}$ (with $q=u,d,s,c$), and
$b\bar{b}\rightarrow Zb\bar{b}$ channels.
To separate the singular regions of phase space and extract
analytically the corresponding IR singularities, we used a
phase-space-slicing (PSS) method with both a soft ($\delta_s$) and a
collinear ($\delta_c$) cutoffs, in terms of which the emission of a
gluon with four-momentum $k$ is defined by the condition
$k^0<\delta_s\sqrt{\hat{s}}/2$ if the gluon is soft, or by the
condition $p_ik<\delta_c k^0 p_i^0$ if the gluon is collinear to
another massless parton with four-momentum $p_i$ (where the momenta
are defined in the parton-level c.m. frame with c.m. energy
$\sqrt{\hat{s}}$).  The procedure is well known and comprehensively
summarized in Ref.~\cite{Harris:2001sx} for both the case of 
initial-state and final-state singularities. In particular, for the 
phase-space integrals originating from a soft emitted parton we used 
the method and expressions reviewed in Ref.~\cite{Denner:1991kt}. 
Results have been cross-checked with two independent codes.

The $O(\alpha)$ real corrections to $bg\rightarrow Zb$ consist of both
photon ($\gamma$) and EW gauge and Higgs boson emissions.  The cross
sections for real $Z/W/H$ emission are finite since their masses
provide a physical IR cutoff and thus can be considered separately.
These processes have very distinct signatures and their inclusion in 
the evaluation of the inclusive cross section for $Z+b$-jet production 
depends on the experimental signature selected.  An example of a set 
of analysis cuts in $Z+j$ production which warrants the inclusion of 
real EW gauge-boson emission in addition to virtual EW corrections
can be found in Ref.~\cite{Baur:2006sn}, where the well-known 
incomplete cancellation of EW Sudakov logarithms between these two
contributions~\cite{Ciafaloni:2000df,Ciafaloni:2000rp} is
  also discussed. Since our study is
not aiming at a detailed analysis of this effect, which should
only be done in collaboration with the experimentalists
performing the measurement of $Z+b$-jet
production, we do not consider the real emission
  of an extra $Z/W/H$. On the other hand, we
consider the QED part of the real radiation
(photon emission) which needs to be consistently included in order to
cancel the IR divergences present in the virtual cross section (due to
photon exchange).

The calculation of the $O(\alpha)$ real-photon emission cross section
can be easily implemented using PSS~\cite{Harris:2001sx}, and actually
involves just a subset of the singularities encountered in the QCD
case for massive $b$ quarks. With due differences, the result can be
easily obtained from there.  On the other hand, the QED case provides
an interesting testing ground for the implementation of the dipole
subtraction (DS) method with initial-state massive dipoles, whose
working knowledge, in the QCD as well as QED cases, is a necessary
step towards the proper implementation of processes with initial-state
massive partons in NLO event generators. All dipoles
for QED radiation off leptons, 
including initial-state massive dipoles, have been calculated
in~\cite{Dittmaier:1999mb}, where they have also been tested against
phase-space slicing in specific QED processes. QED massive dipoles
have also been used and tested against PSS in~\cite{Denner:2009gj} for 
hadronic
$W$+jet production.  In the specific case of photon emission from
the $b$-quark lines of $bg\rightarrow Zb$ we have implemented the
dipole subtraction terms corresponding to the configurations of
massive initial-state
\textit{emitter} and massive final-state \textit{spectator} (both $b$ 
quarks), and vice versa. For illustration, a summary of 
the comparison between the two methods is given in
Table~\ref{tab:pss-vs-ds-photon-emission-mb}, where we report the
total NLO EW cross section calculated using either PSS or
DS. In the case of PSS the result depends on the soft cutoff
$\delta_s$ and we report in
Table~\ref{tab:pss-vs-ds-photon-emission-mb} the values of
$\sigma_{\mbox{\tiny NLO}}^{\mbox{\tiny EW}}$ for
four decreasing values of $\delta_s$, to show the existence of a
plateau region in which $\sigma_{\mbox{\tiny NLO}}^{\mbox{\tiny EW}}$ is actually 
independent of $\delta_s$. We notice that the results reported in
Table~\ref{tab:pss-vs-ds-photon-emission-mb} have been obtained
without imposing a recombination cut, and therefore slightly differ
from what is included in the final results presented in
Section~\ref{sec:results}.  Indeed,
Table~\ref{tab:pss-vs-ds-photon-emission-mb} shows that results
obtained using the PSS method with a soft cutoff $\delta_s$ varying
between $10^{-3}$ and $10^{-5}$ are compatible with results
obtained using the DS method based on
Ref.~\cite{Dittmaier:1999mb}, to which
we refer for more technical details.  If the successful implementation
of the PSS method depends on a careful study of the analytical
dependence on the soft and collinear cutoffs, obtaining meaningful
numerical results using the dipole method involves its own
subtleties. Hence, the cross check between the two methods is all but
trivial, and allows us to move forward to further extensions and
future studies with more confidence.
\begin{table}
\centering
\begin{tabular}{|c|c|c|c|}
\hline
  $\sigma_{\mbox{\tiny LO}}$[pb]  & Method & $\delta_s$ & $\sigma_{\mbox{\tiny NLO}}^{\mbox{\tiny EW}}$[pb] \\\hline
$389.727 \pm 0.005$ & Phase-space slicing & $10^{-2}$  & $383.342 \pm 0.005$ \\
 & & $10^{-3}$  & $383.344 \pm 0.005$ \\
 & & $10^{-4}$  & $383.345 \pm 0.006$  \\
 & & $10^{-5}$  & $383.346 \pm 0.006$  \\
\cline{2-4}
 & Dipole subtraction  &  $-$  & $383.336 \pm 0.006$ \\\hline   
\end{tabular}
\caption{\small Comparison of the NLO EW cross section obtained with
  the PSS and DS method for $Z+b$-jet production at the 13 TeV LHC
  in the setup described in Section~\ref{sec:results}.
  See the text for more details.} 
\label{tab:pss-vs-ds-photon-emission-mb}
\end{table}
We notice that initial-state massive QCD dipoles have first been
studied in Ref.~\cite{Kotko:2012ui}, where however very little detail 
is provided, while a more detailed discussion has recently appeared
in Ref.~\cite{Krauss:2017wmx}.

\subsection{PDF subtraction terms}
\label{sec:pdfsub}

After combining real and virtual NLO QCD or NLO EW corrections,
residual poles in dimensional regularization, due to collinear 
radiation off massless initial-state partons, are absorbed into NLO 
PDF via $\msbar$ factorization. This is systematically achieved by 
defining PDF subtraction terms, also known as collinear counterterms. 

The corresponding treatment in the case of massive initial-state 
quarks needs a more detailed discussion. Indeed, in both NLO QCD and 
NLO EW calculations with a massive $b$ quark, the $b$-quark mass acts 
as a physical regulator for collinear singularities in
$b\rightarrow b(g,\gamma)$ and $(g,\gamma)\rightarrow b\bar{b}$ 
splittings. In particular, radiative corrections involving a massive 
initial-state $b$ quark do not contain collinear poles in dimensional 
regularization, but logarithms of the $b$-quark mass instead, which, 
being finite, do not necessarily need to be absorbed into the
$b$-quark PDF. However, since at LHC energies the mass of the $b$ 
quark is relatively small, these mass logarithms can be large and 
retaining them in the calculation of the partonic cross section, i.e. 
not absorbing them into the $b$-quark PDF, can lead to unnaturally 
large corrections that could eventually affect the numerical stability 
of the hadronic cross section. In a fully massive calculation the 
numerical stability of the hadronic cross section is retained by 
absorbing these mass logarithms via e.g. the generalized $\msbar$ 
scheme of Collins, aka the 
\textit{ACOT scheme}~\cite{Aivazis:1993pi,Collins:1998rz} (see also 
Ref.~\cite{Amundson:2000vg})\,%
\footnote{
  For an overview on the ACOT scheme, as well as its applicability 
  over a range of scales and its convergence to the $\msbar$ scheme in 
  the massless limit, see, e.g., 
  Refs.~\cite{Amundson:2000vg,Tung:2001mv,Dittmaier:2001ay,Kretzer:1998ju}.
  For a review and a comparison of the heavy-quark schemes based on 
  ACOT, TR, and FONLL used respectively in the CTEQ, MSTW (now MMHT) 
  and NNPDF PDF sets, see, e.g., Ref.~\cite{Binoth:2010nha}. A 
  thorough discussion of the case of charm-quark initiated processes 
  in the FONLL scheme can be found for instance in 
  Refs.~\cite{Ball:2015tna, Ball:2015dpa,Ball:2016neh}.
}.
This is achieved by defining corresponding PDF subtraction terms in 
the case of \textit{near-collinear} emission off massive $b$ quarks, 
quite in analogy to subtracting poles in dimensional regularization in 
the case of collinear emission off massless $b$ quarks, such that 
$\msbar$ factorization is retained in the massless limit.

In obtaining the 5FS NLO QCD results presented in
Section~\ref{sec:results} we have used the \textit{S-ACOT scheme} (or
simplified ACOT scheme)~\cite{Kramer:2000hn}, where the $b$ quark is
considered to be massless everywhere, except in the calculation of
those partonic sub-processes that only involve $b$ quarks in the final
state\footnote{As a consequence, both $gg\rightarrow Zb\bar{b}$ and
  $q\bar{q}\rightarrow Zb\bar{b}$ have been calculated with a massive
  $b$ quark in both the 5FS and m5FS case.}.
As such, in this scheme the $b$-quark mass is used as a 
regulator for the $g\rightarrow b\bar{b}$ splitting, while the 
initial-state $b\rightarrow bg$ splitting is treated as for all 
massless quarks (namely, initial-state collinear poles are absorbed in 
the corresponding PDF, by PDF subtraction terms, in $\msbar$ 
factorization). The S-ACOT scheme is what is usually assumed in 
implementing a traditional 5FS (as also done in 
Ref.~\cite{Campbell:2003dd}), and is also what is assumed in defining 
the corresponding PDF in recent CTEQ PDF sets, including the 
\texttt{CT14qed} PDF set that we use for our study\,%
\footnote{
  The \texttt{CT14qed} PDF set~\cite{Schmidt:2015zda} is determined in 
  the same CTEQ global analysis in which the \texttt{CT14} PDF
  set~\cite{Dulat:2015mca} is determined, using the S-ACOT($\chi$) 
  scheme~\cite{Tung:2001mv}, but including QED in the DGLAP evolution 
  of the PDF.
}.

On the other hand, in obtaining NLO QCD and NLO EW results in the m5FS 
we need to implement a fully massive factorization scheme like the 
ACOT scheme, where the $b$ quark is considered massive everywhere. In 
this case, mass logarithms originating from both 
$g\rightarrow b\bar{b}$  (or $\gamma\rightarrow b\bar{b}$, in case 
initial-state photons are considered) and $b\rightarrow bg$ (or 
$b\rightarrow b\gamma$) splittings are absorbed into the $b$-quark 
PDF, and all partonic sub-processes are calculated considering a 
massive $b$ quark\,%
\footnote{
  Although one could use the S-ACOT scheme, which agrees with the ACOT 
  scheme in the high-energy limit, for best accuracy and a fully 
  consistent treatment of massive initial-state quarks, including the 
  possibility of constraining an intrinsic $b$-quark parton density, 
  the use of a scheme like the ACOT scheme should be preferred. For an 
  overview on ACOT vs. S-ACOT see, e.g.,
  Refs.~\cite{Amundson:2000vg,Tung:2001mv,Dittmaier:2001ay,Kretzer:1998ju}.
}.
In accordance with this, in our m5FS calculation of $O(\alpha)$ and 
$O(\alpha_s)$ corrections to $bg\rightarrow Zb$, we subtract from the 
partonic cross section all the mass logarithms arising from both
initial-state $g\rightarrow b\bar{b}$ and $b\rightarrow bg$ (or 
$b\rightarrow b\gamma$) splittings (see Eqs.~\ref{eq:sub_g-bb} 
and~\ref{eq:sub_b-bg}). In our study we use the $b$-quark PDF of the 
\texttt{CT14qed} PDF set~\cite{Schmidt:2015zda}. For the best
accuracy, however, a dedicated PDF set should be determined in the 
ACOT scheme, which is beyond the scope of this study\,%
\footnote{The study presented in Ref.~\cite{Krauss:2017wmx} pursues 
  the direction of determining dedicated PDF sets in a fully massive 
  scheme, also focusing on the definition of a m5FS.
}.

As a reminder on how the PDF subtraction terms enter our calculation,
in a condensed notation, the NLO cross section, using a PSS method, 
can be sketched as follows:
\begin{align}
\sigma
&=\sigma_{\mbox{\tiny LO}}+\sigma_{\mbox{\tiny V}}
+\sigma_{\mbox{\tiny R}}+\sigma_{sub}\\
&=\sigma_{\mbox{\tiny LO}}+\sigma_{\mbox{\tiny V}}
+\Big(\sigma_{\rm soft/coll}\big(\{\delta_s,\delta_c\}\big)
+\sigma_{\rm hard/non-coll}\big(\{\delta_s,\delta_c\}\big)\Big)
+\sigma_{sub}\,,\notag
\end{align}
where $\sigma_{\mbox{\tiny V}}$ denotes the set of UV renormalized
virtual contributions and $\sigma_{\mbox{\tiny R}}$ the set of
real-emission contributions. In the real-emission contributions, soft
and collinear photon/parton emissions are treated using a PSS method
with a soft ($\delta_s$) and a collinear ($\delta_c$) cutoff, as
defined at the beginning of Section~\ref{sec:real}, separating the
real-emission contributions into soft/collinear
($\sigma_{\rm soft/coll}$) and hard/non-collinear
($\sigma_{\rm hard/non-coll}$) contributions.  In the following, we
will collect the expressions of those PDF subtraction terms
($\sigma_{sub}$), which are modified by the presence of a massive $b$
quark or by adopting a m5FS versus a 5FS. Details about the results
reported in this sections can be found in the literature (see, e.g.,
Refs.~\cite{Aivazis:1993pi, Kretzer:1998ju,Baur:1998kt,
  Kramer:2000hn,Dittmaier:2001ay,
  Diener:2005me,Kniehl:2005mk,Harris:2001sx,Amundson:2000vg}).  We
will keep explicit renormalization ($\mu_r$) and factorization
($\mu_f$) dependence, but we will not distinguish between QED and QCD
factorization scales.

As discussed above, in both the 5FS and m5FS scheme, the $b$-quark 
mass appears as physical regulator of the collinear singularity in the 
$g\rightarrow b\bar{b}$ splitting, and the corresponding subtraction 
term reads~\cite{Aivazis:1993pi,Kretzer:1998ju,Amundson:2000vg}
\begin{eqnarray}
\label{eq:sub_g-bb}
\sigma_{sub}^{g \to b\bar{b}}&=&-\int dx_1dx_2\, \frac{\alpha_s(\mu_r)}{2\pi}\,\left\{
\int_{x_1}^1\,\frac{dz}{z}\,
g\left(\frac{x_1}{z},\mu_f\right)g(x_2,\mu_f)\,
\frac{1}{2}\left[z^2+(1-z)^2\right]
\ln\left(\frac{\mu_f^2}{m_b^2}\right)
\, \hat{\sigma}_{\mbox{\tiny LO}} \right.\nonumber\\
&+&(x_1\leftrightarrow x_2)\biggr\}\,,
\end{eqnarray}
where
$\hat\sigma_{\mbox{\tiny LO}}=\hat\sigma_{\mbox{\tiny
    LO}}(x_1,x_2,\mu_r)$
denotes the tree-level partonic cross section for $bg\rightarrow Zb$
(calculated with a massless $b$ quark in the 5FS and with a massive
$b$ quark in the m5FS), and the last line takes into account that an
analogous term where the role of the two gluons is exchanged needs to
be included. Here and in the following it is also understood that the
analogous subtraction terms for $\bar{b}g\rightarrow Z\bar{b}$ are
considered.

In contrast, the collinear $b\rightarrow bg$ (and $b\rightarrow 
b\gamma$) initial-state splittings occurring in the $O(\alpha_s)$
(and $O(\alpha)$) corrections to the $bg$-initiated processes are 
treated quite differently in the traditional 5FS and a m5FS. In a 5FS 
NLO calculation (with a massless initial-state $b$ quark) the 
corresponding PDF subtraction term reads~\cite{Harris:2001sx}
\begin{eqnarray}
\label{eq:sub_b-bg0}
\sigma_{sub}^{b \to b (\gamma,g)}&=&-S_{\eps} \left(\frac{\mu_r^2}{\mu_f^2}\right)^\eps \int dx_1dx_2\left\{
\frac{\alpha_i}{2\pi}\,C_i\,\biggl\{\right.\\
&& \int_{x_1}^{1-\delta_s}\,\frac{dz}{z}\,
b\left(\frac{x_1}{z},\mu_f\right)g(x_2,\mu_f)\,
\left(\frac{1+z^2}{1-z}\right)
\left(-\frac{1}{\epsilon}\right)
\nonumber\\
&+&\left.
b(x_1,\mu_f)g(x_2,\mu_f)\,\left[-\frac{1}{\epsilon} 
\left(2\ln\delta_s+\frac{3}{2}\right)\right] \right\}\, \hat{\sigma}_{\mbox{\tiny LO}} \nonumber\\
&+&(x_1\leftrightarrow x_2)\biggr\}\,,\nonumber
\end{eqnarray}
where $\alpha_i=\alpha$ and $C_i=q_b^2$ (with $q_b$ the charge of the
$b$ quark) for $O(\alpha)$ EW corrections, while $\alpha_i=\alpha_s$
and $C_i=C_F=(N_c^2-1)/2/N_c$ (with $N_c=3$) for $O(\alpha_s)$ QCD
corrections. Here and in the following the last line takes into 
account that an analogous term where $b$ and $g$ come from the 
respectively opposite hadrons needs to be included. Note that, 
performing the $dz$ integration in Eq.~(\ref{eq:sub_b-bg0}), only 
subleading terms in $\delta_s$ remain, which vanish in the limit 
$\delta_s \to 0$ (this is also true for Eqs.~(\ref{eq:sub_b-bg}), 
(\ref{eq:sub_g-gg}) and (\ref{eq:sub_g-gg0}) in the remainder of this 
section). On the other hand, in the m5FS (with a massive initial-state 
$b$ quark) the PDF subtraction term reads~\cite{Kretzer:1998ju,
Baur:1998kt,Dittmaier:2001ay,Diener:2005me,Kniehl:2005mk,
Amundson:2000vg}
\begin{eqnarray}
\label{eq:sub_b-bg}
\sigma_{sub}^{b \to b (\gamma,g)}&=&-\int dx_1dx_2\, \left\{
\frac{\alpha_i}{2\pi}\,C_i\,\biggl\{\right.\\
&&\int_{x_1}^{1-\delta_s}\,\frac{dz}{z}\,
b\left(\frac{x_1}{z},\mu_f\right)g(x_2,\mu_f)\,
\left(\frac{1+z^2}{1-z}\right)
\left[\ln\left(\frac{\mu_f^2}{m_b^2}\frac{1}{(1-z)^2}\right)-1\right]
\nonumber\\
&+&\left.
b(x_1,\mu_f)g(x_2,\mu_f)\,\left[
\ln\left(\frac{\mu_f^2}{m_b^2}\right)\left(2\ln\delta_s+\frac{3}{2}\right)
-2\ln\delta_s-2\ln^2\delta_s+2\right]
\right\}\, \hat{\sigma}_{\mbox{\tiny LO}} \nonumber\\
&+&(x_1\leftrightarrow x_2)\biggr\}\,.\nonumber
\end{eqnarray}

Finally, in the m5FS NLO QCD calculation the subtraction term for the
$g \to gg$ splitting is also affected by the $b$-quark mass as
follows~\cite{Kniehl:2005mk}:
\begin{eqnarray}
\label{eq:sub_g-gg}
\sigma_{sub}^{g \to gg}&=&-S_{\eps} \left(\frac{\mu_r^2}{\mu_f^2}\right)^\eps \int dx_1dx_2\, 
\left\{\frac{\alpha_s(\mu_r)}{2\pi}\,C_g\,\biggl\{\right.\\
&&\int_{x_2}^{1-\delta_s}\,\frac{dz}{z}\,
b(x_1,\mu_f)g\left(\frac{x_2}{z},\mu_f\right)\,
\left[\frac{z}{1-z}+\frac{1-z}{z}+z(1-z)\right]
\left(-\frac{2}{\epsilon}\right)
\nonumber\\
&+&\left.
b(x_1,\mu_f)g(x_2,\mu_f)\,\left[-\frac{1}{\epsilon} 
\left(2\ln\delta_s+\frac{11}{6}-\frac{n_{lf'}}{3C_g}\right)-\frac{1}{9} \ln\left(\frac{\mu_f^2}{m_b^2}\right) \right]
\right\}\, \hat{\sigma}_{\mbox{\tiny LO}} \nonumber\\
&+& (x_1\leftrightarrow x_2)\biggr\}\,,\nonumber
\end{eqnarray}
where $C_g=C_A=N_c$ (with $N_c=3$), and $n_{lf'}=4$ (see also
Section~\ref{sec:virtual_renorm}), whereas in the 5FS the same 
subtraction term reduces to~\cite{Harris:2001sx}
\begin{eqnarray}
\label{eq:sub_g-gg0}
\sigma_{sub}^{g \to gg}&=&-S_{\eps} \left(\frac{\mu_r^2}{\mu_f^2}\right)^\eps \int dx_1dx_2\, 
\left\{\frac{\alpha_s(\mu_r)}{2\pi}\,C_g\biggl\{\right.\\
&&\int_{x_2}^{1-\delta_s}\,\frac{dz}{z}\,
b(x_1,\mu_f)g\left(\frac{x_2}{z},\mu_f\right)\,
\left[\frac{z}{1-z}+\frac{1-z}{z}+z(1-z)\right]
\left(-\frac{2}{\epsilon}\right)
\nonumber\\
&+&\left.
b(x_1,\mu_f)g(x_2,\mu_f)\,\left[-\frac{1}{\epsilon} \left(2\ln\delta_s+\frac{11}{6}-\frac{n_{lf'}}{3C_g}\right)\right]
\right\}\, \hat{\sigma}_{\mbox{\tiny LO}} \nonumber\\
&+&(x_1\leftrightarrow x_2)\biggr\}\,,\nonumber
\end{eqnarray}
with $n_{lf'}=5$. 

\section{Numerical results}
\label{sec:results}

In this section we present results for total cross sections and
distributions of  inclusive $Z+b$-jet production at the LHC with c.m. 
energy 13 TeV. As explained in Sections\,\ref{sec:introduction} 
and~\ref{sec:calculation}, we work in the m5FS and include the first 
order of QCD and EW corrections. We assess their relative impact, and 
the impact of considering $b$-quark mass effects by comparing NLO QCD
results in the 5FS and m5FS.  The values for the SM
input parameters are chosen as follows~\cite{Agashe:2014kda}:
\begin{eqnarray}\label{eq:param}
G_{\mu} = 1.1663787\times 10^{-5} \; {\rm GeV}^{-2},
& \qquad & \alpha(0)= 1/137.035999074, \quad \alpha_s\equiv\alpha_s(M_Z^2)=0.118
\nonumber \\
M_Z = 91.1876 \; {\rm GeV}, & \quad & M_W = 80.385 \; {\rm GeV},  \quad 
M_H = 125 \; {\rm GeV}
\nonumber  \\
m_e  = 0.510998928 \; {\rm MeV}, &\quad &m_{\mu}=0.1056583715 \; {\rm GeV},
\quad m_{\tau}=1.77682 \; {\rm GeV}
\nonumber  \\
m_u=0.06983 \; {\rm GeV}, & \quad & m_c=1.275 \; {\rm GeV},
\quad m_t=173 \; {\rm GeV}
\nonumber  \\
m_d=0.06983 \; {\rm GeV}, & \quad & m_s=0.15 \; {\rm GeV}, 
\quad m_b=4.75 \; {\rm GeV}
\end{eqnarray}
The weak mixing angle is calculated from the weak gauge-boson masses,
i.e. $\sin^2\theta_w=1-M_W^2/M_Z^2$.  The lepton and light-quark
masses are only used in the evaluation of the renormalization constant
for the electric charge, $\delta Z_e$, resulting into large
logarithmic corrections.  Such large universal EW corrections can be
absorbed into the corresponding lowest-order cross sections by using
specific EW input schemes, i.e. the $\alpha(M_Z)$-scheme and the
$G_\mu$-scheme as described, e.g., in ~\cite{Hollik:1988ii,
  Denner:1991kt,Dittmaier:2001ay,Andersen:2014efa}.  Here we provide
results in the so-called $G_\mu$-scheme according to which in the
Born-level couplings $\alpha(0)$ is replaced by $\alpha_{G_\mu}$:
\begin{equation}\label{eq:alphagmu}
\alpha(0) \to \alpha_{G_\mu}=\frac{\sqrt{2}G_\mu M_W^2}{\pi} 
(1-\frac{M_W^2}{M_Z^2})\,\,\,,
\end{equation}
and $\delta Z_e$ receives a contribution from
$\Delta r$~\cite{Sirlin:1980nh}, which describes the EW one-loop
corrections to muon decay.  For the input parameters given in
Eq.~(\ref{eq:param}) we find $\Delta r=0.02968$.  Consequently, the
NLO EW cross section in the $G_\mu$-scheme is related to the NLO EW
cross section in the $\alpha(0)$-scheme as follows:
\begin{equation}\label{eq:gmu}
\sigma_{\mbox{\tiny NLO}}^{\mbox{\tiny EW},G_\mu}=
(\sigma_{\mbox{\tiny NLO}}^{\mbox{\tiny EW},\alpha_0}-
\Delta r \, \alpha(0) \, \alpha_s \, \sigma_{\mathrm{LO}}^{(1,1)}) \frac{\alpha_{G_\mu}}{\alpha(0)}\,\,\,.
\end{equation}
In this way, the large logarithmic dependence on the lepton and light
quark masses due to $\delta Z_e$ in the EW one-loop corrections is
completely canceled by the corresponding contribution in $\Delta r$,
while EW universal corrections are included and resummed in the Born
cross section.  Note that we choose the relative EW corrections to be
evaluated with $\alpha(0)$, i.e. in the virtual and real EW
corrections $\alpha(0)^2$ is replaced by $\alpha(0) \alpha_{G_\mu}$,
which is appropriate for photonic corrections. Alternatively, one can
evaluate the relative corrections using $\alpha_{G_\mu}$, which
differs from our choice by higher-order corrections.

\begin{figure}
\centering
\includegraphics[scale=0.625]{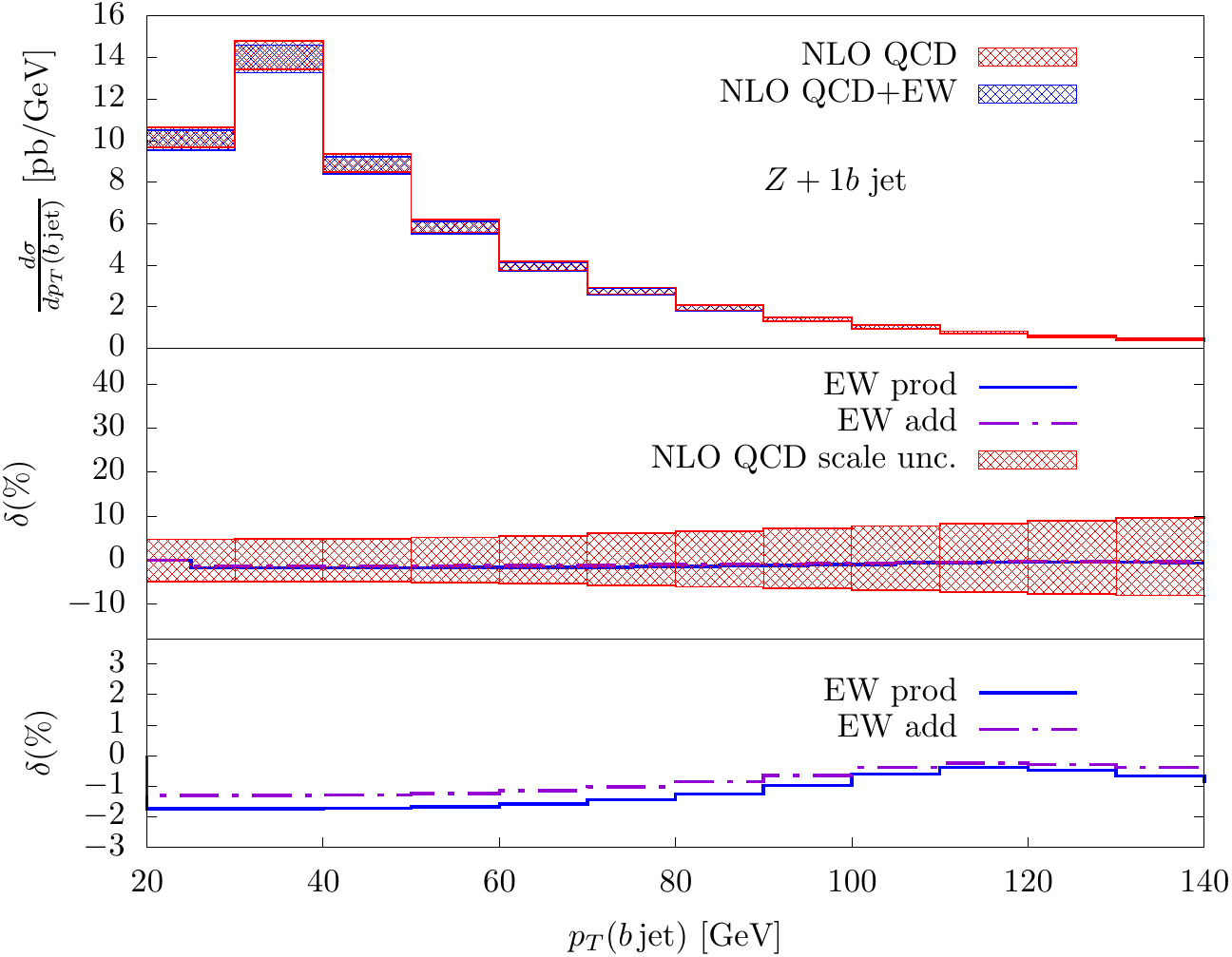}%
\includegraphics[scale=0.625]{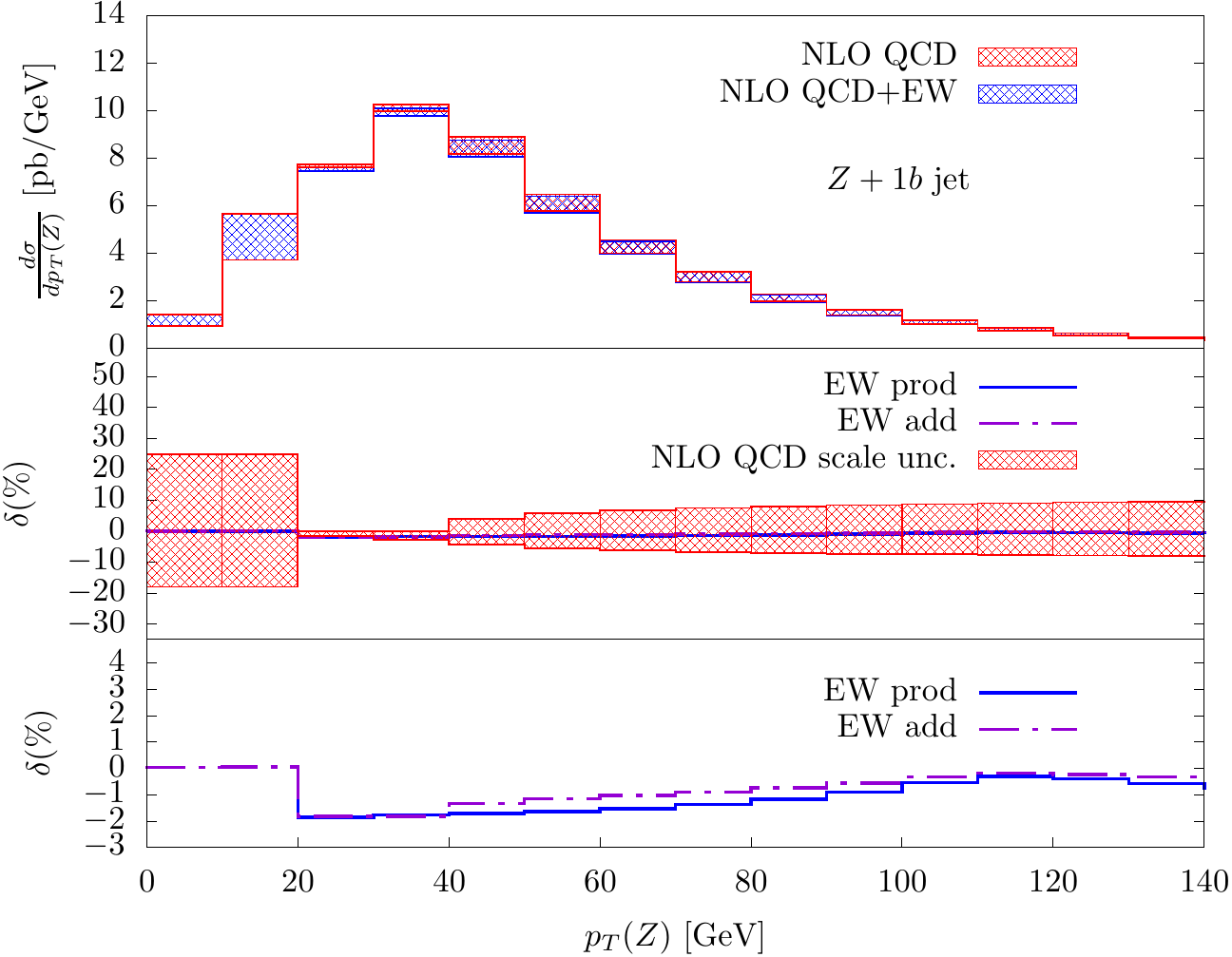}\\
\includegraphics[scale=0.625]{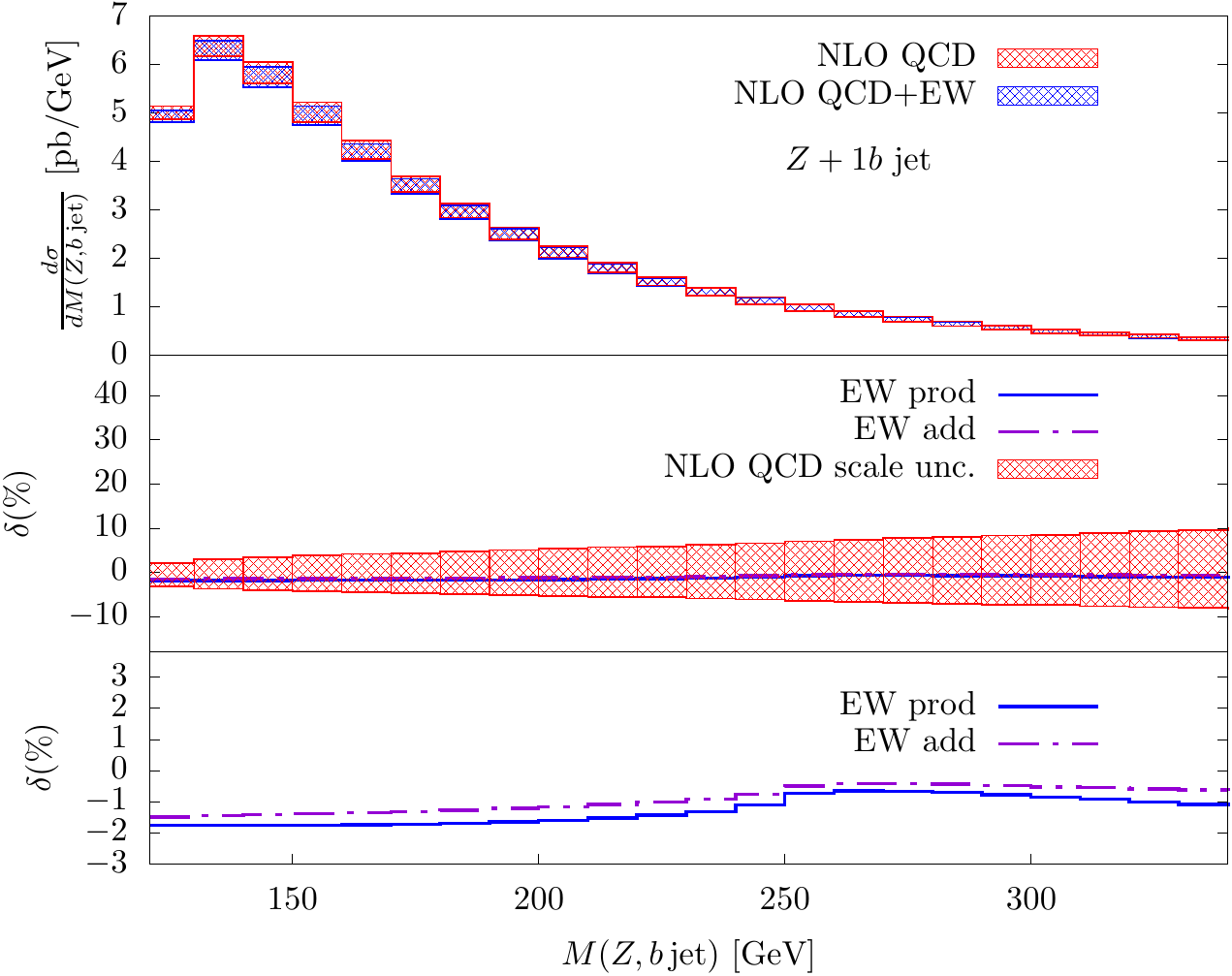}
\caption{\small Differential distributions for the $b$-jet and $Z$
  transverse momentum, and the invariant mass of the $Z$ boson and the $b$-jet.
The lower plots show the relative EW $O(\alpha)$ corrections differential
distributions, $\delta_{\mbox{\tiny EW}}^{\mbox{\tiny prod}}$ and
$\delta_{\mbox{\tiny EW}}^{\mbox{\tiny add}}$, together with the NLO QCD scale
uncertainty in the middle plot.}
\label{fig:dsigmadpt-Z-b}
\end{figure}

\begin{figure}
\centering
\includegraphics[scale=0.625]{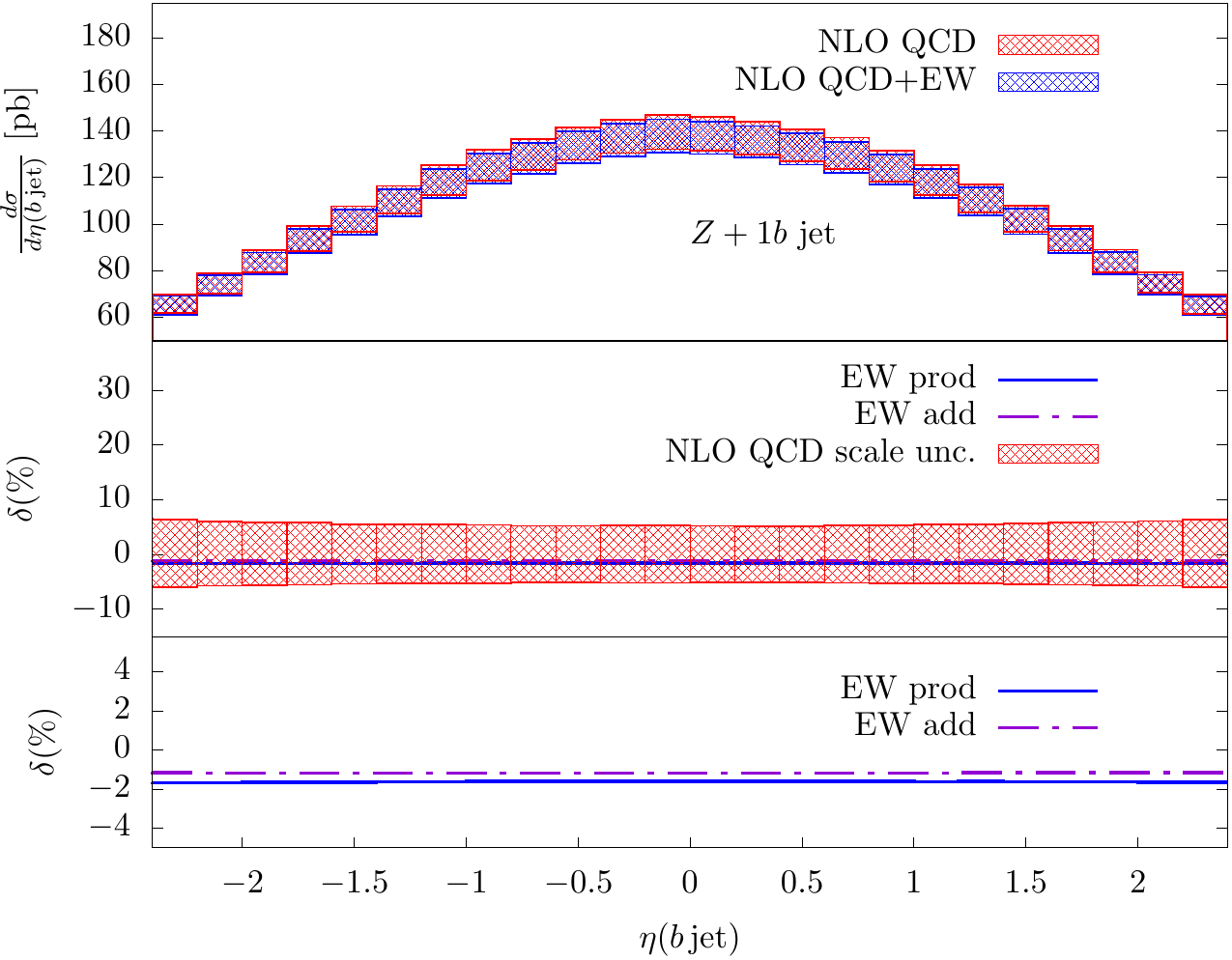}
\includegraphics[scale=0.625]{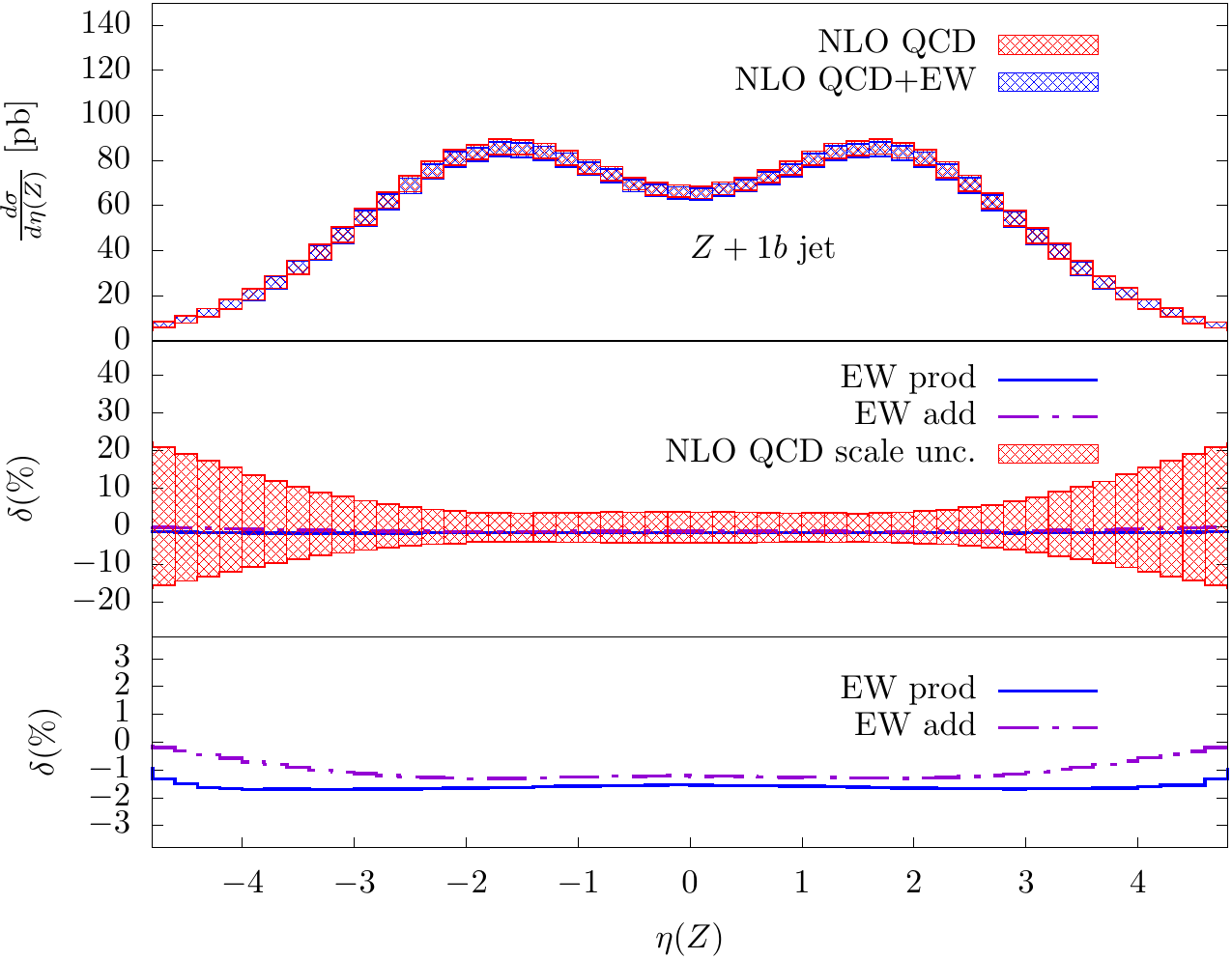}
\caption{\small Differential distributions for the $Z$ boson and $b$-jet pseudo rapidity.
The lower plots show the relative EW $O(\alpha)$ corrections differential
distributions, $\delta_{\mbox{\tiny EW}}^{\mbox{\tiny prod}}$ and
$\delta_{\mbox{\tiny EW}}^{\mbox{\tiny add}}$, together with the NLO QCD scale
uncertainty in the middle plot.}
\label{fig:dsigmadeta-Z-b}
\end{figure}

\begin{figure}
\centering
\includegraphics[scale=0.6]{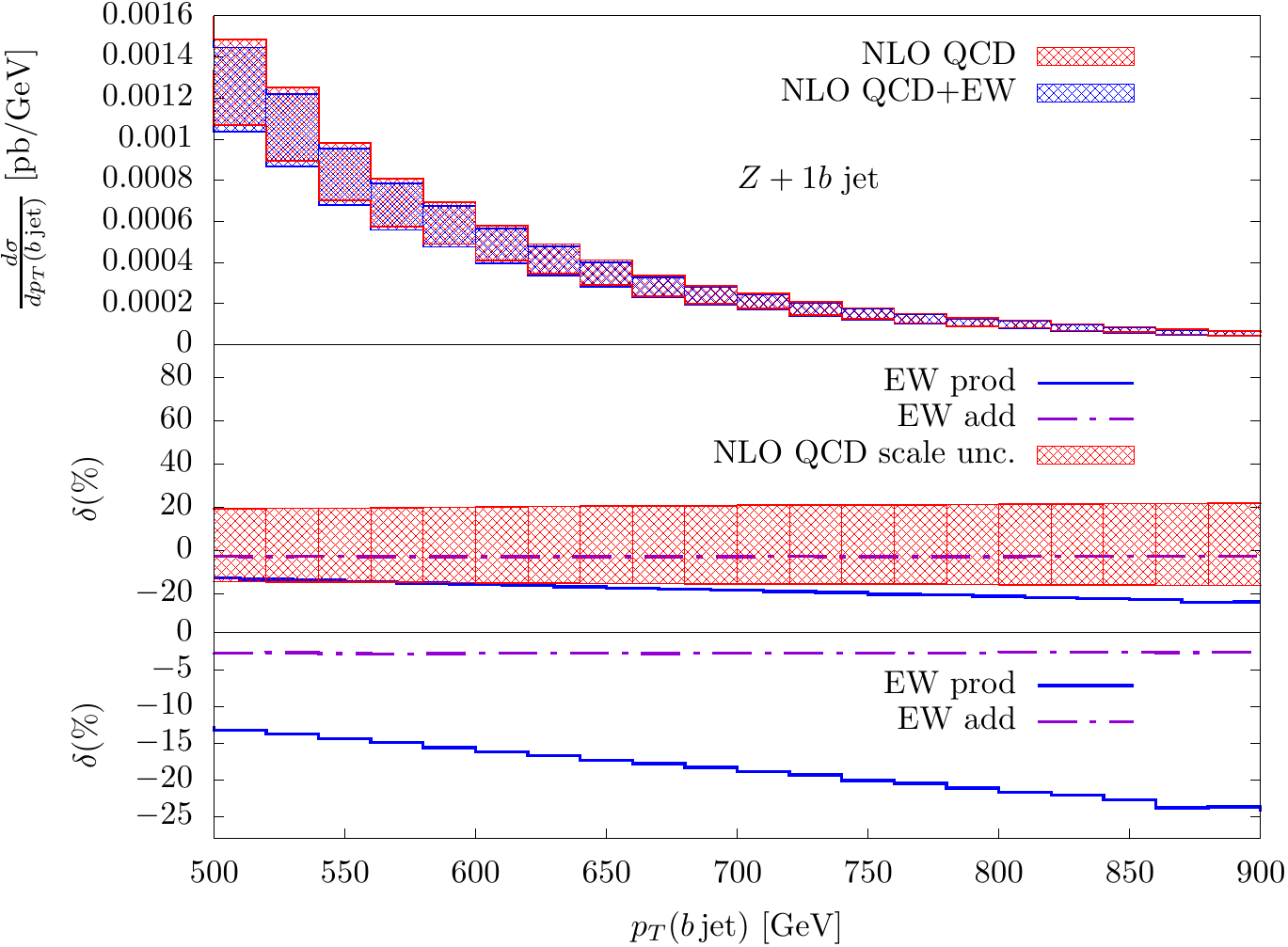}%
\includegraphics[scale=0.6]{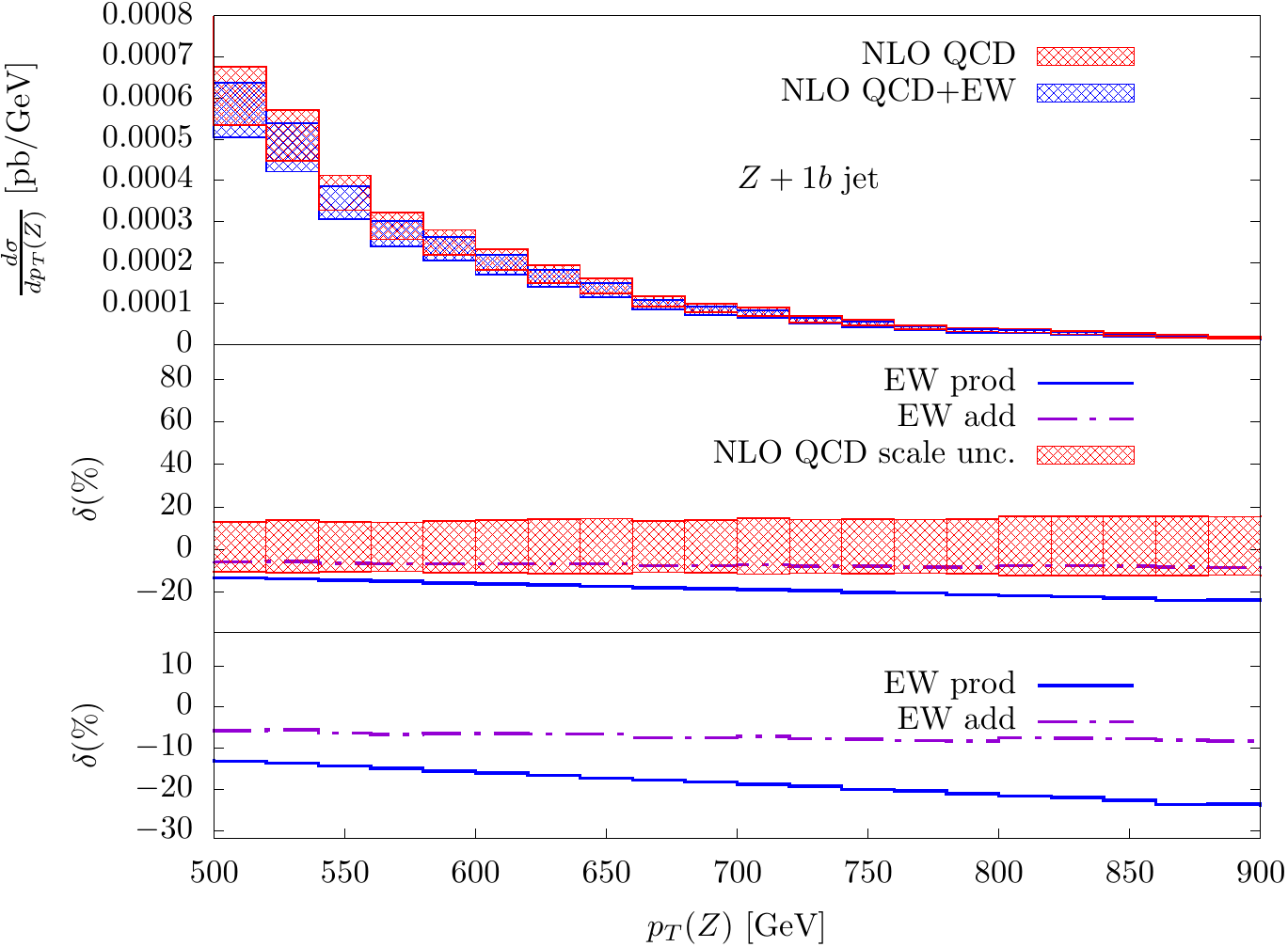}\\
\includegraphics[scale=0.6]{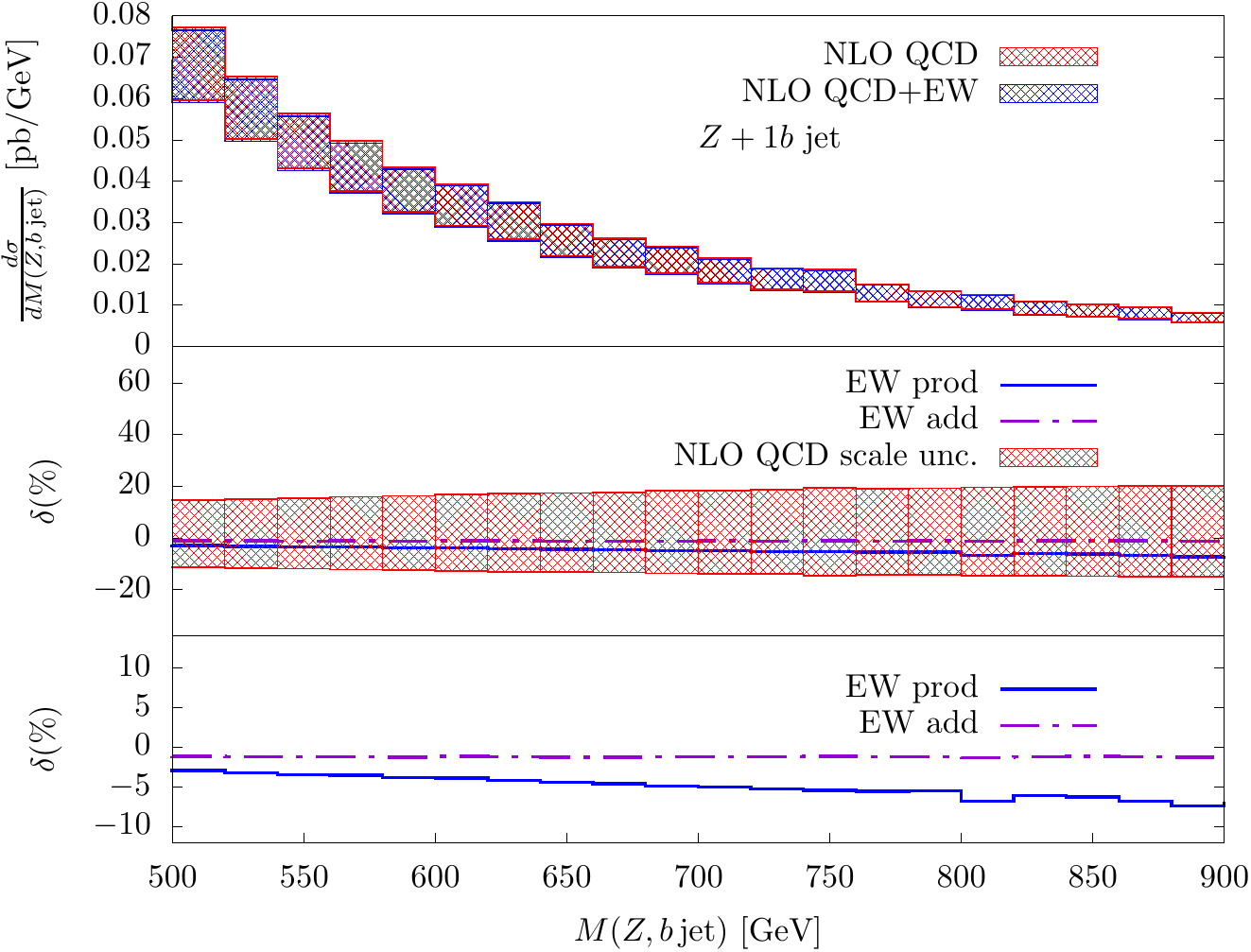}
\caption{\small Differential distributions for the $b$-jet and $Z$
  transverse momentum, and the invariant mass of the $Z$ boson and the $b$-jet at
high $p_T$ and high invariant $Z\,b$-jet mass. The lower plots show the relative
EW $O(\alpha)$ corrections differential distributions,
$\delta_{\mbox{\tiny EW}}^{\mbox{\tiny prod}}$ and
$\delta_{\mbox{\tiny EW}}^{\mbox{\tiny add}}$, together with the NLO QCD scale
uncertainty in the middle plot.}
\label{fig:dsigmadpt-Z-b-high}
\end{figure}

\begin{figure}
\centering
\includegraphics[scale=0.625]{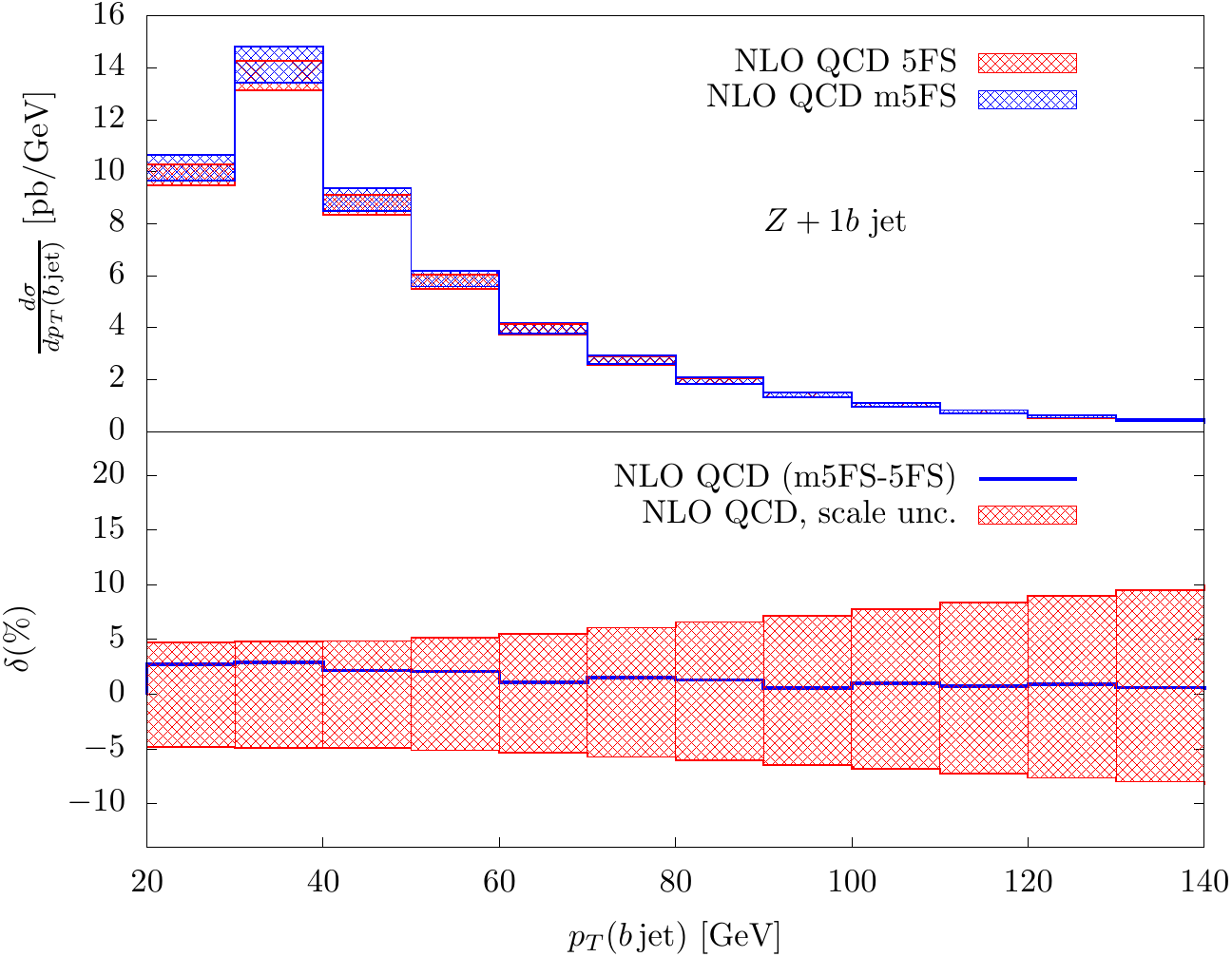}%
\includegraphics[scale=0.625]{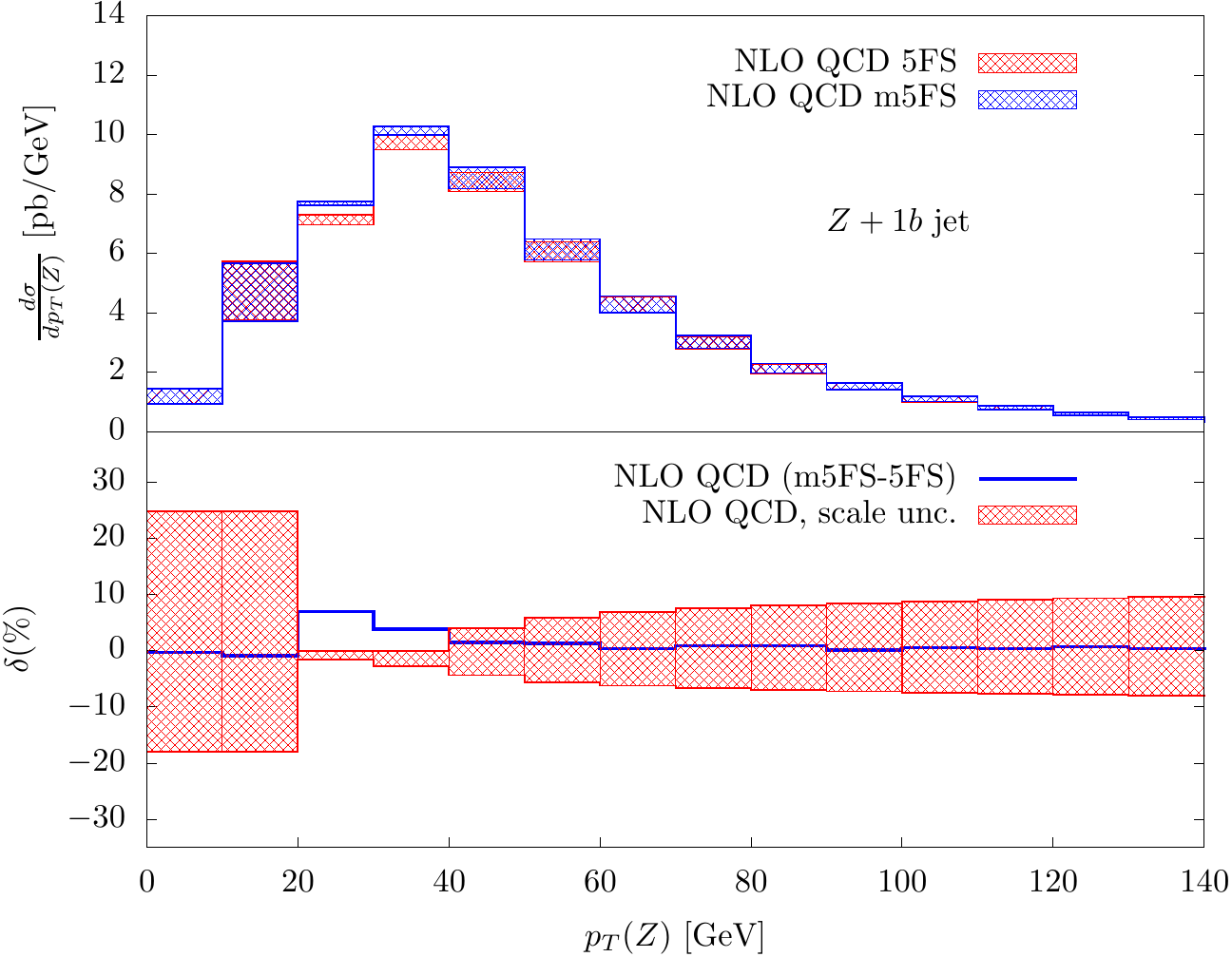}\\
\includegraphics[scale=0.625]{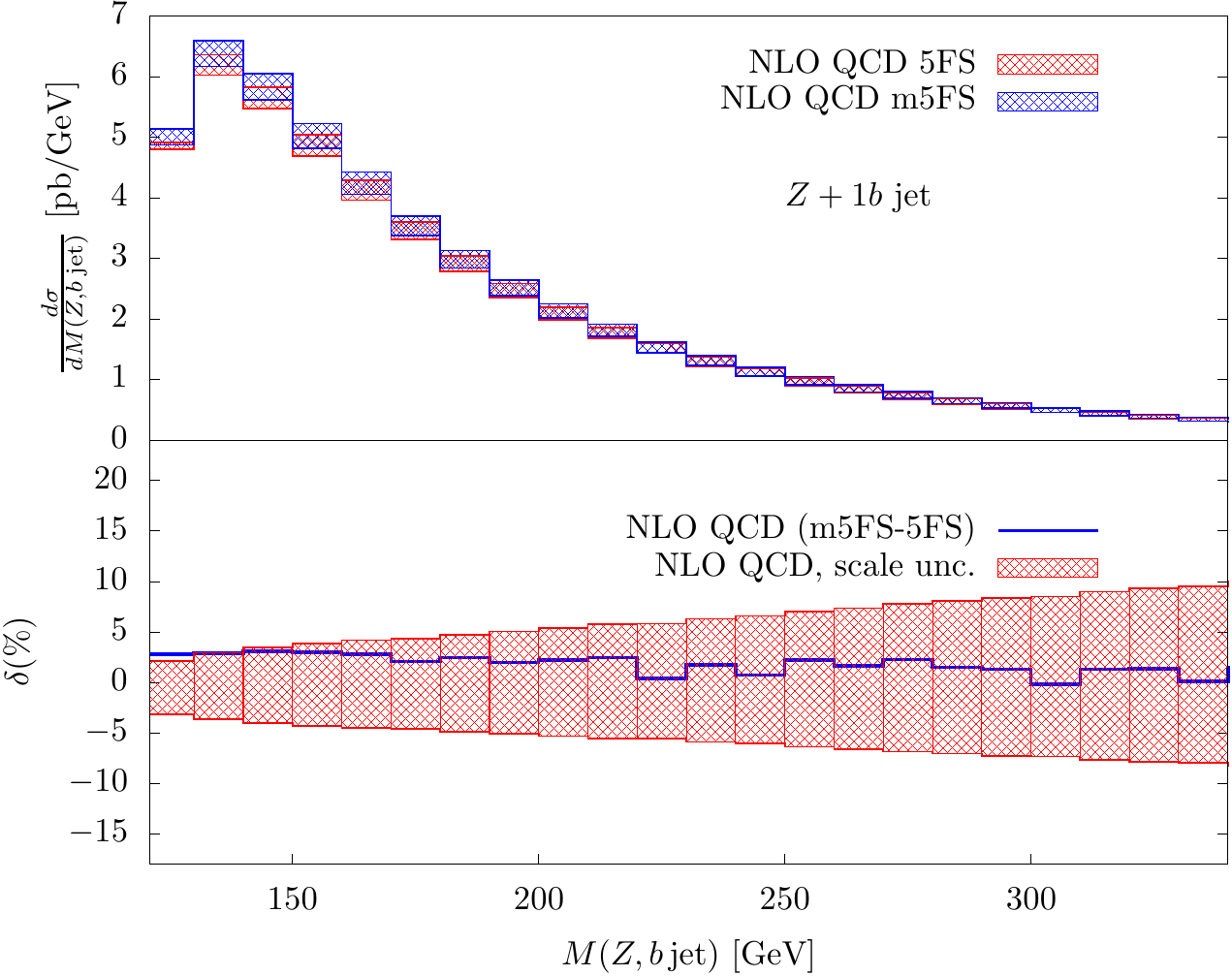}
\caption{\small 5FS and m5FS NLO QCD predictions
  for differential distributions for the $b$-jet and $Z$ transverse momentum,
  and the invariant $Z\,b$-jet mass. The lower plot shows the difference
$\delta_{\mbox{\tiny mb}}$ together with the NLO QCD scale uncertainty.}
\label{fig:dsigmadpt-Z-b-qcd}
\end{figure}

\begin{figure}
\centering
\includegraphics[scale=0.625]{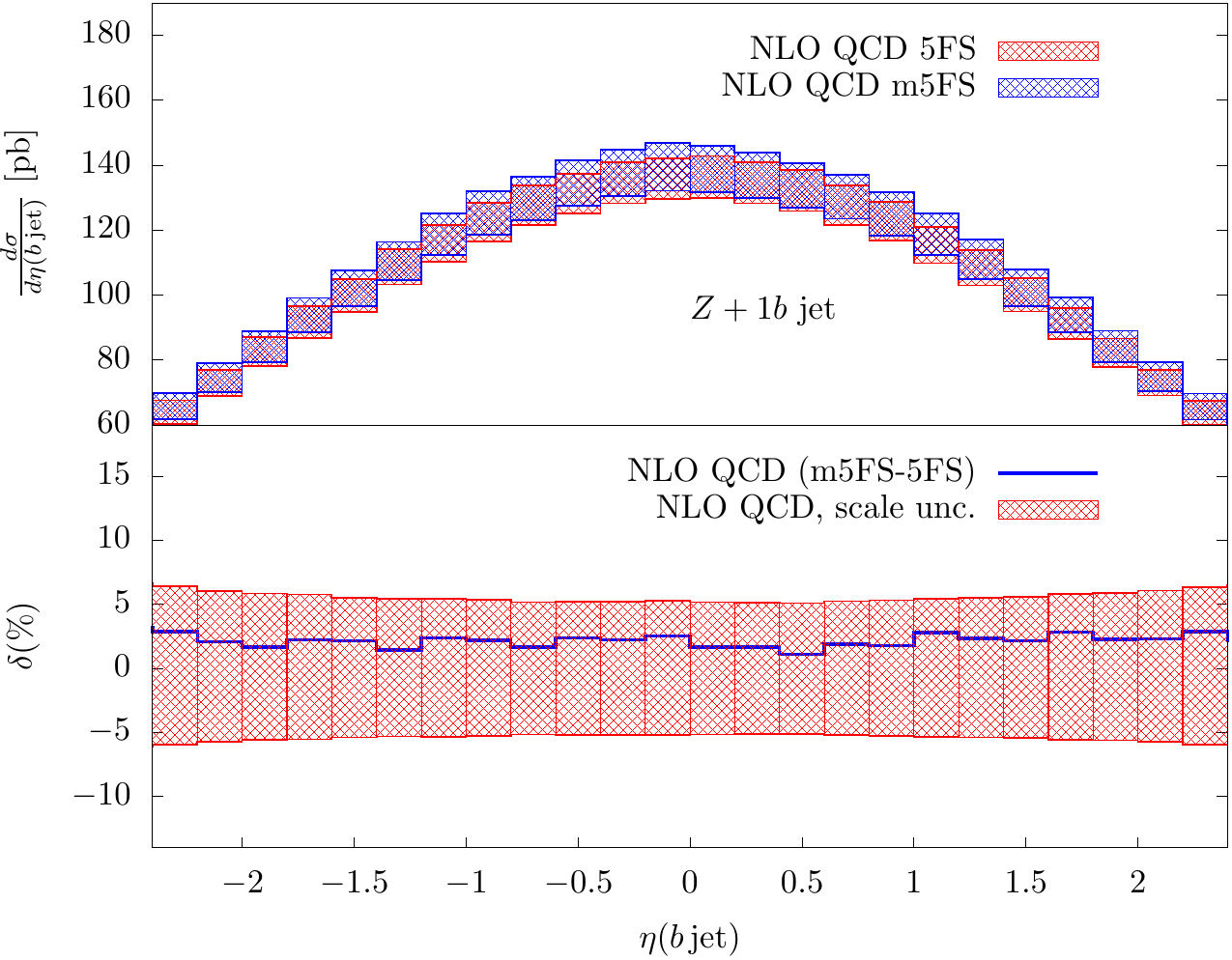}
\includegraphics[scale=0.625]{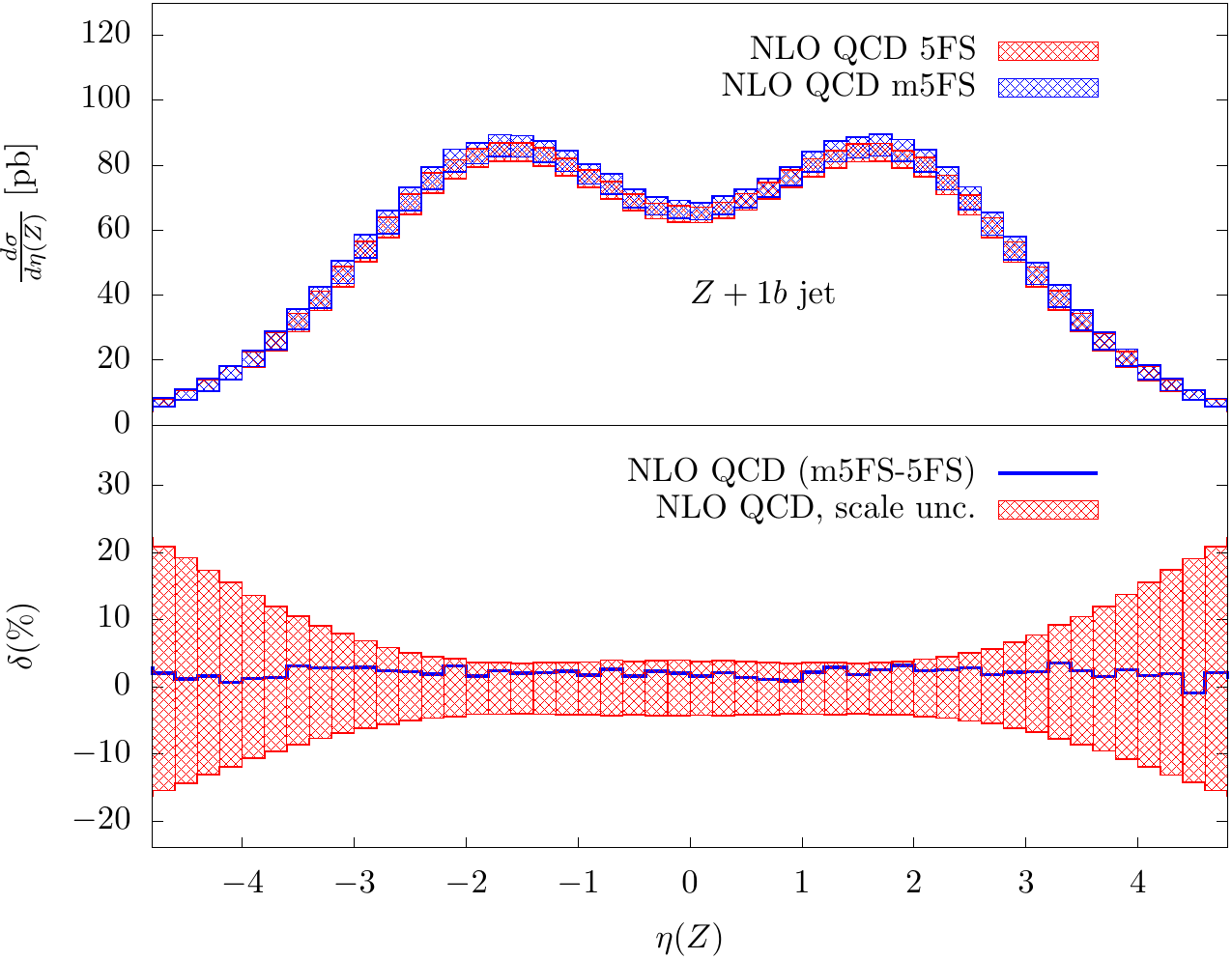}
\caption{\small 5FS and m5FS NLO QCD predictions for
  differential distributions for the $Z$ and $b$-jet pseudo-rapidity. The lower
plot shows the difference, $\delta_{\mbox{\tiny mb}}$, together with the NLO QCD
scale uncertainty.}
\label{fig:dsigmadeta-Z-b-qcd}
\end{figure}

Since we include both QCD and QED corrections, we choose to use the
\texttt{CT14qed} PDF set~\cite{Schmidt:2015zda} with
$\alpha_s(M_Z)=0.118$. We do not include in our study an estimate of
the current PDF uncertainty, since, for $b$-quark initiated processes,
this will become more meaningful when a complete implementation of the
m5FS is included in the available PDF sets. Likewise, we do not
consider uncertainties from varying the $b$-quark mass. On the other
hand, the choice of a PDF set which includes QED radiation is
necessary when considering EW corrections. In both NLO and LO results
we use NLO PDF and the two-loop running of $\alpha_s$ with $n_{lf}=5$
flavors (c.f. Eq.~(\ref{eq:dZgs})). We choose to quantify the
uncertainty from scale variation by varying $\mu_r$ by a factor of two
above and below the central value $\mu_r=M_Z$, while keeping $\mu_f$
fixed at $\mu_f=M_Z$.  Our choice is motivated by the fact that most
of the residual scale dependence is driven by the NLO QCD cross
section which varies monotonically with $\mu_r$ and $\mu_f$ in
opposite directions~\cite{Campbell:2003dd}. It is therefore more
appropriate not to vary the two scales together setting
$\mu_r=\mu_f$. At the same time, varying both scales simultaneously
and independently is not necessarily a better estimate of the
theoretical uncertainty, which to be more accurate would require a
dedicated study of both fixed and dynamical scale choices as well as
other factors, like the choice of different PDF sets. We notice
however that a study of the $p_T$ distributions of leading and
subleading $b$ jets in $pp\rightarrow Zb\bar{b}$ including NLO QCD
corrections~\cite{Cordero:2009kv} seems to favor a choice of $\mu_f$
anywhere between $50-100$~GeV (where the typical $dp_T/p_T$ rescaling
responsible for the leading part of the integrated bottom PDF is more
evident). Hence our choice of $\mu_f=M_Z$. The bands presented in
Figs.~\ref{fig:dsigmadpt-Z-b}-\ref{fig:dsigmadeta-Z-b-qcd} have been
built using this prescription.

Since we consider inclusive $Z+b$-jet production, we include $Z+b+X$
and $Z+\bar{b}+X$ final states, with
$X=\{\mbox{light parton}, \gamma, b\, (\mbox{or}\,\bar{b})\}$, in the
real-emission case.  As we have at most two partons in the final
state, we use a simple jet (recombination) algorithm,
i.e. we recombine the final-state $b$ (or $\bar{b}$) quark with a
final-state light quark, gluon, photon, $\bar{b}$ (or $b$) quark (for
$Zb\bar{b}$ events), if their separation in the azimuthal angle-pseudo
rapidity plane,
$\Delta R(b,X)=\sqrt{(\Phi_b-\Phi_X)^2+(\eta_b-\eta_X)^2}$, is less
than $R_{\mbox{\tiny min}}=0.4$, as would be the case for any cone
algorithm of this size.
Moreover, we impose the following
acceptance cuts on the transverse momentum and pseudo-rapidity of all
$b$ jets: $p_T(b) > 25$~GeV and $|\eta(b)| < 2.5$, and we keep all
events that have at least one $b$ jet. For events with 2 $b$ jets, the
differential distributions for $b$-jet observables show the
hardest $b$ jet.
 
\begin{table}[H]
\centering
\begin{tabular}{|c|c|c|c|c|c|}
\hline
&LO & NLO QCD & NLO EW & NLO QCD+EW  & NLO QCD $\times$ EW\\ \hline \hline   
$\sigma$(pb) & 
$389.73^{+46}_{-37}(392.66^{+46}_{-37})$ & 
$537.7^{+30}_{-29}(526.9^{+29}_{-28})$ &
$383.40^{+44}_{-36}$ & 
$531.4^{+30}_{-29}$  & $529.2^{+30}_{-29}$ \\ \hline
$\delta$(\%) &
- & 38(34) & -1.6 & 36 & 36 \\\hline \hline
\end{tabular}
\caption{\small Total cross sections at LO and NLO (first row) including only QCD NLO
  corrections, only EW NLO corrections, or both (in the
  \textit{additive} and \textit{multiplicative} approach), at the LHC
  with c.m. energy 13 TeV. The results obtained with $m_b=0$ at LO and
  NLO QCD are given in  parenthesis. 
  The central values are obtained for $\mu_r=\mu_f =M_Z$ and the
  quoted uncertainties are calculated varying $\mu_r$ by a factor of
  two about the central value, for fixed $\mu_f=M_Z$.
  The relative impact of each NLO contribution,
  $\delta(\%)=(\sigma_X/\sigma_{\mbox{\tiny LO}}-1)\times 100$
  (for $X$=NLO QCD, NLO EW, $\ldots$), is given in
  the second row. See text for more details. 
  The errors reported in sub- and superscripts are purely from scale 
  variation, while the statistical error (not reported) is on the last digit of
  the given results.}
\label{tab:results}
\end{table}
In Table~\ref{tab:results} we present results for the total cross
sections. In order to illustrate the impact of different orders of
corrections on the total cross section we give results for both LO and
higher-order cross sections, and we distinguish between the
NLO cross sections obtained by including only QCD corrections
($\sigma_{\mathrm{NLO}}^{\mathrm{QCD}}$), only EW
corrections ($\sigma_{\mathrm{NLO}}^{\mathrm{EW}}$), or both. We have
combined NLO QCD and EW corrections according to both the
\textit{additive} ($\sigma_{\mathrm{NLO}}^{\mathrm{QCD}+EW)}$) and
\textit{multiplicative} ($\sigma_{\mathrm{NLO}}^{\mathrm{QCD}\times EW)}$)
approaches. While in the \textit{additive} approach the two sets of
corrections are simply added 
(c.f. Eq.~(\ref{eq:sigma_nlo})), i.e.
\begin{align}
\sigma_{\mbox{\tiny NLO}}^{\mbox{\tiny QCD+EW}}
=\sigma_{\mbox{\tiny LO}}
+\alpha_s^2\alpha\,\sigma^{(2,1)}
+\alpha_s\alpha^2\sigma^{(1,2)}\,,
\end{align}
while in the \textit{multiplicative} approach the
NLO QCD result is multiplied by the EW $K$-factor
($K_{\mbox{\tiny EW}}$)~\cite{Campbell:2016dks}:
\begin{equation}
\sigma_{\mbox{\tiny NLO}}^{\mbox{\tiny QCD}\times \mbox{\tiny EW}}=
\sigma_{\mbox{\tiny NLO}}^{\mbox{\tiny QCD}} \times
\frac{\sigma_{\mbox{\tiny NLO}}^{\mbox{\tiny EW}}}{\sigma_{\mbox{\tiny LO}}} 
\equiv \sigma_{\mbox{\tiny NLO}}^{\mbox{\tiny QCD}} \times K_{\mbox{\tiny EW}}\,,
\end{equation}
hence including mixed QCD$\times$EW corrections that are of higher
order.  As a first estimate of the overall effect of each kind of NLO
corrections and their combined effect on the LO cross
section, in the second row of Table~\ref{tab:results} we give the
relative corrections as percentage of the LO cross section,
$\delta(\%)=(\sigma_X/\sigma_{\mbox{\tiny LO}}-1)\times 100$
(where X=NLO QCD, NLO EW, $\ldots$).

If the impact of NLO EW corrections on the total cross section is
indicative of the average magnitude of their effect, a much more
interesting result is their effect on distributions. Furthermore, it
is important to compare the effect of NLO EW corrections to the
residual theoretical uncertainty of NLO distributions, including both
QCD and EW corrections, and estimate in particular whether NLO EW
corrections are within the scale uncertainty of the corresponding
NLO QCD corrections.  With this in mind, in the following we will
quantify the impact of EW corrections in terms of the following ratios:
\begin{equation}
\label{eq:delta-ew-add}
\delta_{\mbox{\tiny EW}}^{\mbox{\tiny add}}=\frac{\sigma_{\mbox{\tiny NLO}}^{\mbox{\tiny QCD+EW}}-
\sigma_{\mbox{\tiny NLO}}^{\mbox{\tiny QCD}}}{\sigma_{\mbox{\tiny NLO}}^{\mbox{\tiny QCD}}}
=\frac{\sigma^{\mbox{\tiny QCD+EW}}_{\mbox{\tiny NLO}}}{\sigma^{\mbox{\tiny QCD}}_{\mbox{\tiny NLO}}}-1\,,
\end{equation}
and
\begin{equation}
\label{eq:delta-ew-prod}
\delta_{\mbox{\tiny EW}}^{\mbox{\tiny prod}}=\frac{\sigma_{\mbox{\tiny
      NLO}}^{\mbox{\tiny QCD}\times \mbox{\tiny EW}}-
\sigma_{\mbox{\tiny NLO}}^{\mbox{\tiny QCD}}}{\sigma_{\mbox{\tiny NLO}}^{\mbox{\tiny QCD}}}
=\frac{\sigma^{\mbox{\tiny EW}}_{\mbox{\tiny NLO}}}{\sigma_{\mbox{\tiny LO}}}-1\,,
\end{equation}
which correspond to the \textit{additive} and \textit{multiplicative}
approach of combining NLO QCD and EW cross sections respectively.

Figs.~\ref{fig:dsigmadpt-Z-b} to~\ref{fig:dsigmadpt-Z-b-high}
illustrate the results of our calculation in terms of several
distributions. In Fig.~\ref{fig:dsigmadpt-Z-b} we present the
differential distributions for the transverse momentum of the
final-state $b$-jet ($p_T(b$ jet)) and $Z$ boson ($p_T(Z)$) as well as
their invariant mass ($M(Z,b\,\rm jet)$) in the region of low $p_T$
and low invariant mass, while the analogous distributions in the
region of high $p_T$ and high invariant mass are given in
Fig.~\ref{fig:dsigmadpt-Z-b-high}. Given the large difference between
the magnitude of the corresponding distributions at low and high
momenta, we separate the two regions to illustrate the effects of EW
corrections both at the peak (low-$p_T$/mass region) and in the tail
(high-$p_T$/mass region) of such distributions.  The upper plots of
these figures show the comparison between the NLO distributions
obtained including only QCD corrections or both QCD and EW
corrections. Each distribution is given as a band obtained by
considering the renormalization-scale variation for fixed $\mu_f$, as
explained earlier. For the sake of readability, in the upper plots we
only compare to the \textit{additive} combination of QCD and EW NLO
corrections, while we consider both cases in the lower plots.  Indeed,
the lower plots of these figures illustrate both
$\delta_{\mbox{\tiny EW}}^{\mbox{\tiny add}}$ and
$\delta_{\mbox{\tiny EW}}^{\mbox{\tiny prod}}$, as defined in
Eqs.~(\ref{eq:delta-ew-add}) and (\ref{eq:delta-ew-prod}), and compare
them to the bin-by-bin uncertainty of the NLO QCD cross section. The
same information is provided in Fig.~\ref{fig:dsigmadeta-Z-b} for the
$b$-jet and $Z$-boson pseudorapidity distributions ($\eta(b\,\rm jet)$
and $\eta(Z)$).

The results presented in Figs.~\ref{fig:dsigmadpt-Z-b}
to~\ref{fig:dsigmadpt-Z-b-high} confirm that the effect of EW
corrections both on $p_T$ and $\eta$ distributions are at the level of
a few percent as for the total cross section (see
Table~\ref{tab:results}) , apart from the high $p_T$ regions.  The
impact of the EW $O(\alpha)$ corrections on the LO cross sections (see
Eq.~(\ref{eq:delta-ew-prod})) can be seen in the relative corrections
of the multiplicatively combined QCD and EW NLO corrections in the
lower plots of Fig.~\ref{fig:dsigmadpt-Z-b-high}, in particular at
high $p_T$, where they reduce the central-scale LO cross section by
$\sim$ 24\% at $p_T=900$ GeV. However, as discussed in
Section~\ref{sec:real} these effects will have to be re-examined when
experimental analyses require the inclusion of real EW gauge-boson
radiation.  In the low-$p_T$ region the impact of EW corrections is
completely within the scale uncertainty of the differential NLO QCD
cross section\footnote{We notice that while the $p_T^b$ spectrum is
  cut at low $p_T$ by a tagging cut, the $p_T^Z$ spectrum at NLO can
  extend to vanishing $p_T$. The unusually large scale dependence in
  the region around the $b$-jet $p_T$ cut is due to well known
  instabilities in matching the LO and NLO phase spaces.} while at
high $p_T$ EW corrections may be large enough to exceed the QCD
uncertainty. For analyses focusing on the high-$p_T$ region this
statement will have to be confirmed by dedicated studies of the
theoretical uncertainty in a realistic simulations that not only
included the effect of scale variation. In the case of pseudorapidity
distributions, both the additive and multiplicative combination of QCD
and EW NLO corrections show that the effect of EW corrections is still
hidden in the NLO QCD uncertainty affecting these distributions, with
the central $\eta(Z)$ region being the least affected.

Following the same logic, we quantify the effect of switching from a
massless 5FS to a m5FS in terms of the ratio
\begin{equation}
\delta_{\mbox{\tiny mb}}=\frac{\sigma_{\mbox{\tiny NLO}}^{\mbox{\tiny QCD,m5FS}}-
\sigma_{\mbox{\tiny NLO}}^{\mbox{\tiny QCD,5FS}}}{\sigma_{\mbox{\tiny NLO}}^{\mbox{\tiny QCD,5FS}}}
=\frac{\sigma_{\mbox{\tiny NLO}}^{\mbox{\tiny QCD,m5FS}}}{\sigma_{\mbox{\tiny NLO}}^{\mbox{\tiny QCD,5FS}}}-1\,,
\end{equation}
representing the fractional change in the NLO QCD cross sections in
going from the 5FS to the m5FS.  This ratio is given in the lower
plots of Figs.~\ref{fig:dsigmadpt-Z-b-qcd} and
\ref{fig:dsigmadeta-Z-b-qcd}, where we present results for the
final-state $b$-jet and $Z$-boson $p_T$ distributions, and their
invariant-mass and pseudorapidity distributions. In these figures, the
upper plots give the explicit form of the distributions in the 5FS and
m5FS, together with their bin-by-bin scale uncertainty. Most of the
observed mass effects still lie within the uncertainty of the
differential NLO QCD cross sections, in particular for $p_T$ and
invariant-mass distributions. The largest deviations are at the level
of a few percent and are concentrated as expected in the
low-$p_T$/mass region. On the other hand the pseudorapidity
distributions are sensitive to mass effects over their entire
range. We do indeed expect that a modification of the initial-state
kinematics can lead to modifications of the final-state angular
distributions. This corroborates our original motivation for a
consistent implementation of a m5FS in Monte Carlo event generators.

Finally, it is clear from our study that a better control of
higher-order QCD corrections is still a limiting factor in achieving
further theoretical accuracy in the prediction of $Z+b$-jet
production.  The residual still sizable scale dependence of both total
and differential cross sections in the m5FS (and 5FS) does not come as
a surprise since in the calculation of the $O(\alpha_s^2\alpha)$ part
of the NLO cross section for $bg\rightarrow Zb$ new important channels
(such as $gg\rightarrow Zb\bar{b}$) appear for the first time, and
introduce a large Born-like scale dependence in the NLO QCD cross
section.  As already emphasized in Section~\ref{sec:introduction},
this large scale dependence could be greatly reduced by calculating
the NNLO QCD corrections to $bg\rightarrow Zb$, where the known NLO
QCD corrections to $gg\rightarrow Zb\bar{b}$~\cite{Cordero:2009kv}
contribute, possibly retaining the $b$ quark initial-state mass
dependence.

\section{Conclusions}
\label{sec:conclusions}

In this paper we have considered the production of a $Z$ boson with at
least one $b$ jet as a phenomenologically interesting testing ground
to assess the impact of finite initial-state $b$-quark masses, and to
evaluate the relevance of a still missing piece of NLO corrections,
the combined QCD and EW NLO corrections. We have consistently
implemented the initial-state massive kinematics and the corresponding
matching with existing PDF, setting the stage for a consistent
development of a massive 5FS, which should provide the correct
interface between fixed-order NLO calculations and parton-shower Monte
Carlo event generators.  We have presented the first calculation of
the first-order EW corrections to $bg\rightarrow Zb$, using the \nlox
one-loop provider, and implementing real photon-emission corrections
using both a phase-space slicing method and dipole subtraction with
massive initial-state
dipoles. EW and QCD NLO corrections have been combined using both an
additive and multiplicative approach.

Both mass effects and NLO EW corrections are small effects, compared
to the size of NLO QCD corrections, and are mostly within the
uncertainty of the NLO QCD cross section. Still, there are clear
indications of their impact, in particular on the entire spectrum of
angular distributions, and in complementary transverse-momentum
regions: EW corrections mainly affect the high-$p_T$ region of both
$b$-jet and $Z$ boson $p_T$, while $b$-quark mass effects are more
pronounced in the low-$p_T$ regions.

Given the phenomenological relevance of $Z+b$-jet production, both as
background to Higgs-boson precision measurements and new physics
searches, as well as a potential candidate for a direct precision
measurement of the $b$-quark PDF, we should aim to reduce the
theoretical uncertainty in the future. In this respect, it is clear
that the inclusion of NNLO QCD corrections will greatly help to
mitigate unphysical scale dependencies. At the same time, the
consistent inclusion of $b$-quark mass effects in existing PDF sets,
where the $b$-quark mass only plays the role of an IR regulator so
far, can only improve our control of the Monte-Carlo generation of $b$
initiated processes.

\section*{Acknowledgements}
We thank Pavel Nadolsky, Fred Olness, and Dave Soper for
discussions. L.~R. and D.~F. would like to particularly thank Fernando
Febres-Cordero for his interest in clarifying some technical aspects
of the implementation of massive initial-state dipoles.  The work of
S.~H., S.~Q., L.~R., and C.~R. had been and is supported in part by
the U.S. Department of Energy under grant DE‐SC0010102.  The work
of D.~W. is supported in part by the U.S. National Science Foundation
under Grant No. NSF-PHY-1417317.  S.~H., L.~R., C.~R., and D.~W. are
grateful for the hospitality of the Kavli Institute for Theoretical
Physics (KITP) during the workshop on \textit{LHC Run II and the
  Precision Frontier} at which part of this work was being prepared.
Their research at the KITP was supported in part by the National
Science Foundation under Grant No. NSF PHY-1748958.  L.~R. would like
to also thank the Aspen Center for Physics for the hospitality offered
while parts of this work were being completed.

\bibliographystyle{apsrev}
\bibliography{zbew.bbl}

\end{document}